\documentclass{article}

\usepackage{amssymb,amsfonts,amsmath}
\usepackage{graphicx}

\providecommand\bnabla{\boldsymbol{\nabla}}

\newsavebox{\astrutbox}
\sbox{\astrutbox}{\rule[-5pt]{0pt}{20pt}}

\begin{document}
\title{Numerical simulations of an effective two-dimensional model 
for flows with a transverse magnetic field}


\author{A.Poth\'erat, J.Sommeria and R Moreau\\
$^{1}$ Ilmenau Technical University, Faculty of Electrical Engineering, \\
Kirchhoffbau, Kirchhoffstrasse 1, 98693 Ilmenau, Germany\\
$^{2}$LEGI (CNRS),
ENSHMG BP 95 38402 Saint Martin d'H\`{e}res Cedex,  France \\
$^{3}$Laboratoire EPM-MADYLAM (CNRS)\\
ENSHMG BP 95 38402 Saint Martin d'H\`{e}res Cedex, France}

\date{30 September, 2004}








\maketitle



\begin{abstract}
This paper presents simulations of the 2d model developed by \cite{psm00} for MHD
flows between two planes with a strong transverse
homogeneous and steady magnetic field, accounting for
moderate inertial effects in Hartmann layers. We first show analytically how the
additional terms in the equations of motion accounting for inertia,
soften velocity gradients in the horizontal plane, and then we implement
the model on a code to carry out numerical simulations to be
compared with available experimental results. This comparison shows that
the new model can give very accurate results as long as the Hartmann layer
remains laminar. Both experimental velocity profiles and global angular
momentum measurements are closely recovered, and local and global Ekman
recirculations are shown to alter significantly the aspect of the flow as well 
as the global dissipation. 
\end{abstract}

\section{Introduction}

The velocity field in liquid metal flows under a strong magnetic field tends to 
vary very little along the magnetic field lines so that in many situations, such flows are almost 
two-dimensional. This striking property of this particular kind of MHD flow  
was first studied in  the 70's (\cite{kolesnikov74}) and can be observed in many laboratory experiments 
and industrial applications (\cite{buhl96}). For instance, it can drastically modify heat and mass transfer 
in the liquid metal blankets used in Tokamak-type nuclear fusion reactors. These blankets carry
a liquid metal confined between two planes, and are submitted to a typical 
$10$ T magnetic field, 
required to confine hot plasma inside the reactor. Their role is to evacuate the heat generated by
nuclear fusion within the plasma and to regenerate the tritium which feeds the reaction itself.
The efficiency of the whole device is therefore tightly bound with the properties of the 
quasi-2d  turbulent flow which takes place within the blankets.\\

The fact that the velocity is almost uniform along the magnetic field lines, 
except in the vicinity of walls non parallel to the field where thin boundary 
layers develop (Hartmann layers), provides interesting perspectives 
for modelling. It is indeed
tempting to derive a simplified effective 2d equation for the outer velocity from the full 3d equations.
This is achieved by averaging the full Navier-Stokes equations along the  
direction of the magnetic field, which yields a 2d model. 
The advantages of this approach are numerous: firstly, it saves a significant amount of
 computational resources as the 3d problem is replaced by a 2d one. Secondly,
 when the boundary layer is thin, the analytical treatment in a 
2d model may be more accurate than a 3d numerical solution that cannot adequately resolve
the boundary layer (See \cite{tagawa02}).
Finally, this approach is general, because these models solely rely on 
assumptions on the values of non-dimensional numbers and include no empirical assumption or 
 empirical parameter.  It is also a general approach in the sense that 2d models involve 
 no assumptions on the 
component of the flow perpendicular to the direction the magnetic field
(it can be turbulent for instance).\\
This approach itself is not new and has already been successfully used in MHD 
(\cite{verron87, buhl96, psm00}) for flows confined between parallel planes. It 
had been used even before this to model rotating fluid layers such as oceans and 
atmospheres (see for instance \cite{green69,pedlosky87}). Flows dominated
 by a strong rotation are indeed analogous to MHD flows in the sense that the 
 velocity also varies little along the rotation vector, except in the vicinity 
 of walls where Ekman boundary layers develop.\\

	The physical problem  of particular interest in this paper, is that of MHD flows 
confined between two parallel horizontal plates and plunged in a strong, vertical,
 steady and uniform magnetic field $\mathbf B$. The flow is driven by injection of current at 
 one of the plates. The references to horizontal and vertical directions are for 
 ease of description as gravity has no relevance here. 
 This problem exhibits all the features of the quasi-2d flows described 
above. It is of interest in industrial applications (nuclear fusion reactor blankets
 as well as continuous casting of steel processes) and in laboratory experiments. 
 In most of these situations, the magnetic Reynolds number $Rm$
 is small so that the change in $\mathbf B$ due to the currents induced by the flow
is $O(Rm)$ and may be neglected.
 In such cases, \cite{sm82}
 have shown that electromagnetic effects reduce to a diffusion of momentum along 
 the magnetic field lines. If this phenomenon is stronger than inertial effects 
 (\textit{i.e.} the interaction parameter $N$, which represents the ratio of 
 electromagnetic and inertial forces is greater than unity ) and 
 viscous effects (\textit{i.e.} the Hartmann number $Ha$, the square of  which represents the ratio of electromagnetic and viscous forces, is greater than unity), then the 
flow is 2d, except in the vicinity of walls non-parallel to the magnetic field
where viscosity balances 
 electromagnetic effects to give rise to the Hartmann boundary layer
 (see for instance \cite{moreau90}). \cite{sm82} have derived a 2d model 
 (denoted SM82 thereafter) based 
  on the simple exponential profile of Hartmann layers. It gives good results
 in problems where inertia is small (see \cite{psm00} and \cite{delannoy99}) 
 but fails to describe flows where some strong rotation gives rise to 3d secondary flows,
 such as Ekman pumping. \cite{psm00} have developed a 2d model accounting for such 
 phenomena (denoted PSM2000 thereafter). We shall here use both models in order to explain the 
 results of two MHD experiments which have not been modelled up to now: \cite{sommeria88}'
 s electrically driven vortices and the MATUR experiment. The example of PSM2000 
 emphasise that 2d models can be highly refined to account for rather complex 3d flows, 
 whilst still retaining the advantages of working in 2d. This underlines the 
 flexibility of 2d models.\\

	The layout of the paper is as follows: in section 2, we briefly summarise the 
principles of 2d models and describe SM82 and PSM2000. We also show that the 
effects of local 3d recirculations accounted for in the latter is to smooth the
 vorticity field. In section 3, we describe how the models are implemented by a numerical
 code and perform a convergence test under grid refinement to test the reliability of 
  the whole system. In section 4, PSM2000 is used to recover experimental results on
the free decay of isolated vortices of \cite{sommeria88}.
 Section 5 is devoted to the study of the complex flow involved in the MATUR 
 experiment developed in Grenoble. In particular, we show how local and global 
 recirculations re-shape the flow, firstly by the spectacular smoothing effect 
 theoretically described in section \ref{sec:properties}, and secondly \textit{via} 
 the additional dissipation induced by the thinning of the boundary layers formed on 
 the vertical side walls which confine the flow.

\section{2d models and properties}
\label{sec:properties}
\subsection{General configuration and averaged equations}

\begin{figure}
\begin{center}
\includegraphics[width=0.9\textwidth]{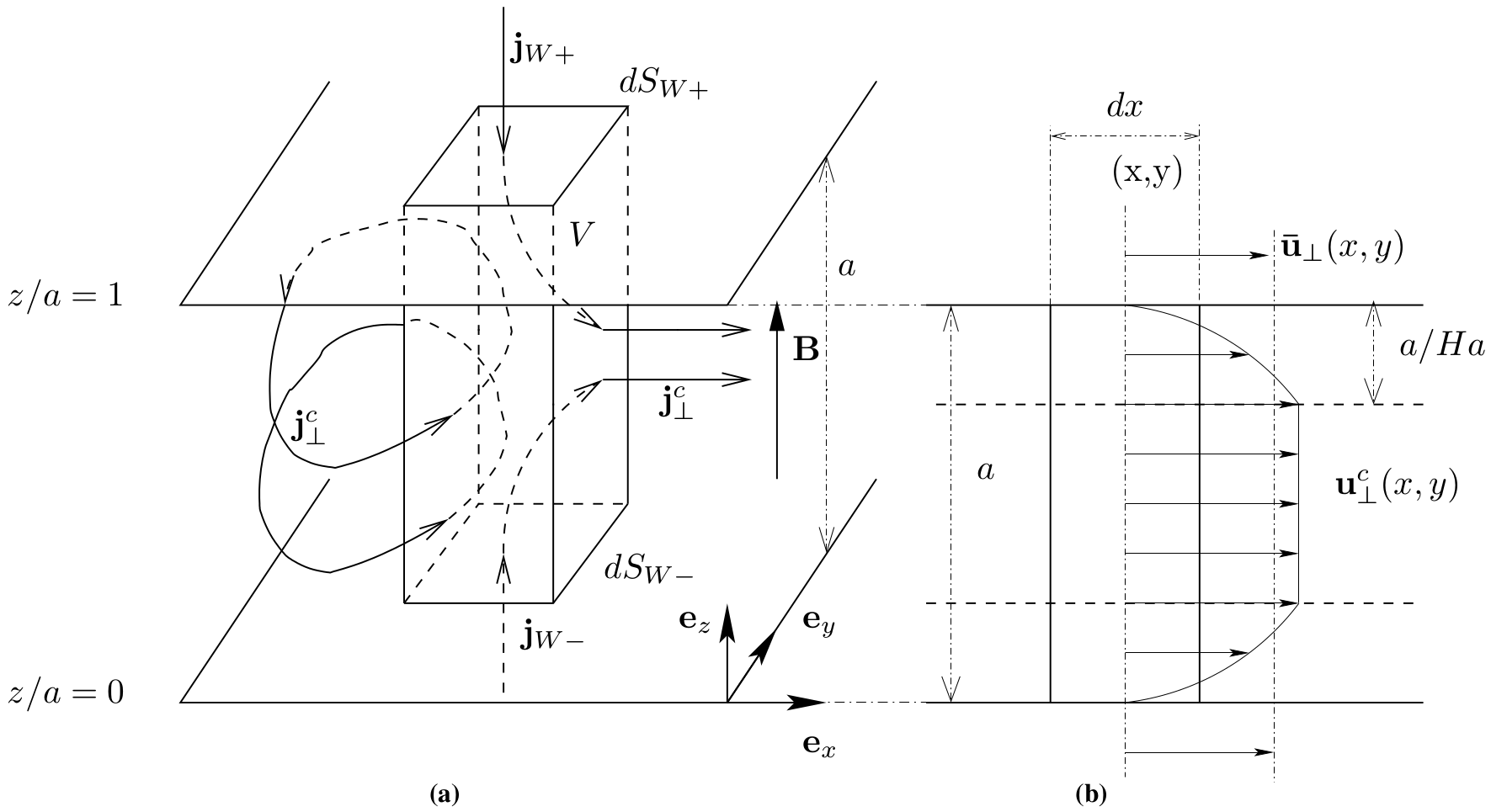}
\caption{General flow configuration,
with control surface V to which momentum conservation is applied in order 
to derive the general equation for 2d models (\ref{2d integrate motion equation})}
\label{2dmod}
\end{center}
\end{figure}

A fluid of density $\rho $, kinematic viscosity $\nu $ and electrical
conductivity $\sigma $ is  assumed to flow between two parallel electrically
insulating plates (spacing $a$) orthogonal to the uniform magnetic field $%
\mathbf{B}$. As explained above, we state that $\mathbf{B}$ is vertical for simplicity of
description but there is no gravity effect. For strong enough magnetic
fields, the velocity is independent of the vertical
coordinate $z$, except in the thin Hartmann layers (thickness $aHa^{-1}$)
located on the horizontal plates. The velocity in the core (\textit{i.e.}
outside of these layers) is then close to the averaged velocity between $z=0$
and $z=1$ to a precision of $Ha^{-1}$ (lengths are normalised by $a$).
A good model of the dynamics is then obtained by averaging the horizontal 
components of the Navier-Stokes equations between the two plates.
The starting point of such a 2d model is the momentum equation for the 
control volume illustrated in figure \ref{2dmod}. Its cross-sectional area
(in planes $z=const$) is uniform but of 
infinitesimal size.
%
%
Rewriting the equation derived by \cite{psm00} in non-dimensional terms (normalisation by 
fluid depth $a$, typical velocity $U$, time $a/U$, pressure $\rho U^2$, 
shear stress $(\rho \sigma U/a) Ha$ and electric current density $\sigma B U/Ha$), we get
\footnote{A typical distance in the direction
perpendicular to the field $l_\perp=a/ \lambda$ was necessary for other aspects of the work 
presented in \cite{psm00}. In the present paper, all distances are normalised by $a$, which
is equivalent to choosing $\lambda=1$. We have also chosen a scaling which brings the friction 
to the leading order. This differs from  the original scaling in \cite{psm00} but reflects the 
physics of the SM82 and PSM2000 models more accurately.}:
\begin{equation}
\partial _{t}\mathbf{\bar{u}}_{\bot }+
\mathbf{\bar{u}}_{\bot }.\bnabla_\bot \mathbf{\bar{u}}_{\bot }
+\mathbf{\nabla }_\bot\bar{p} 
-\dfrac{N}{Ha^2}\mathbf{\nabla}^2_{\bot }\mathbf{\bar{u}}_{\bot }
-\frac{N}{Ha}\left(\mathbf{\bar{j}}_\bot\times\mathbf e_z\right)=
-\overline{\left( \mathbf{u}^{\prime }\mathbf{.\nabla }\right)\mathbf{u}^{\prime }}
+\frac{N}{Ha}\mathbf{\tau }_{W},
\label{2d integrate motion equation}
\end{equation}%
%
where the over-bar denotes z-averaging across the fluid depth ($z=0$ to $z=1$). $\mathbf u^\prime$ represents the departure 
from the averaged velocity from the average $\mathbf{ \bar u}$, so that the 
average of $\mathbf u^\prime$ is zero.
Quantities averaged along $z$  are by definition dependent only on
$x$ and $y$. The corresponding Nabla operator $\bnabla_\bot$ is 
two dimensional and carries the subscript ()$_{\bot }$. Similarly, the same 
subscript on 
a vector indicates components perpendicular to the magnetic field only.\\

The two important non-dimensional numbers mentioned in 
section 1 appear: the 
Hartmann number $Ha=aB\sqrt{\sigma/\rho\nu}$, and the interaction parameter 
$N=\sigma B^2 a/(\rho U)$.

The $z$-average of the 
$\mathbf{u}_{\bot }\mathbf{.\nabla}_\bot \mathbf{u}_{\bot }$ does not 
reduce to 
$\mathbf{\bar{u}}_{\bot }\mathbf{.\nabla}_\bot \mathbf{\bar{u}}_{\bot }$:
like in turbulence, a "Reynolds stress" 
$\overline{\left( \mathbf{u}^{\prime }\mathbf{.\nabla }\right)\mathbf{u}^{\prime }}$
appears, involving the deviation $\mathbf{u}^{\prime }$ from the averaged 
velocity. The
first term on the right hand side is effectively 
a Reynolds-stress term arising from the departure to the average of the 
velocity along the field direction $\mathbf e_z$. The non-dimensional wall 
stress term $\mathbf \tau_W$ is the average
of stresses on the planes at $z=0$ and at $z=1$, and is dependent on the $(x,y)$ coordinates only.\\
%
At low $Rm$, the Ohm's law is linear. The equations governing continuity of electric current and  incompressibility 
are also linear so they may be averaged to give:
\begin{eqnarray}
\begin{array}{cc}
\bnabla.\mathbf{\bar j}_\bot=-j_W, &
\bnabla.\mathbf{\bar u}_\bot=0 \\
\end{array}
\end{eqnarray}
\begin{equation}
\frac{1}{Ha}\mathbf{\bar j}= \mathbf{\bar E}+\mathbf{\bar u}\times \mathbf e_z
\label{eq:ohm}
\end{equation}
where $j_W$ is the current density injected at one or both of the confining planes
and $\mathbf E$ is a non-dimensional electric field. Taking the curl of the 
Ohm's law and using the incompressibility condition, one sees that $\mathbf{\bar j_\bot}$ is
irrotational. It follows that there is a potential $\psi_0$ for 
$\mathbf{\bar j}_\bot$ which satisfies Poisson's equation, the source term being
$j_W$:
\begin{equation}
\begin{array}{cc}
\mathbf{\bar j}_{\bot }=\mathbf{\nabla }_\bot\psi _{0}, & 
\bnabla_{\bot}^2\psi _{0}=-j_{W}.%
\end{array}
\label{current}
\end{equation}
The potential $\psi_0$ is determined from the current source as the solution of this Poisson equation (\ref{current}), which is unique for a given current flux
$\mathbf{\bar j}_{\bot }.\mathbf n$ at the lateral boundaries. Then, 
using the vector field $\mathbf u_0$ of streamfunction $\psi_0$, the
 Lorentz force in equation (\ref{2d integrate motion equation}) turns out to 
 depend on the boundary condition on
 the electric current as $\mathbf{\bar j}_\bot \times  \mathbf e_z=\mathbf u_0$.\\
 
%
 At this point,
we insist that no approximation has been made on the equations of motion.
The next step is to then express $\mathbf \tau_W$ and the Reynolds-stress tensor using physical
models derived from asymptotic expansions performed  on the full 3d equations of motion
equation. We give two examples of the  resulting 2d models in the next two paragraphs,
which are going to be used to perform numerical simulations throughout
 the rest of this paper. For more detail about the derivation of these models,
 the reader is referred to \cite{psm00}.

\subsection{The SM82 model}

\cite{sm82} were the first to construct a 2d model based on the above ideas. 
They used the classical Hartmann layer profile for the boundary layer model
and assumed that the velocity and pressure in the core do not depend on $z$
(the 2d\ core model). These two assumptions are of first order in the limits 
$N\rightarrow \infty$ and $Ha\rightarrow \infty$, keeping the ratio $Ha/N$ 
finite (\textit{i.e.} assuming that $Ha$ and $N$ are
of comparable orders of magnitude).
The Hartmann layer theory states that $\mathbf \tau_W$ is related to the excess current 
in the Hartmann layer with:
\begin{equation}
\mathbf \tau_W=n(\mathbf{\bar j}-\mathbf j^c_\perp)\times \mathbf e_z 
=n(\mathbf u^c_\perp \times \mathbf e_z)\times \mathbf e_z=-n\mathbf u^c_\perp.
\end{equation}
This relates $\mathbf \tau_W$ to the core electric current $\mathbf j^c$ and velocity $\mathbf u^c$. 
$n$ is the number of Hartmann layers in the flow. $n=1$ if the upper
plane $z=1$ is a free surface, and $n=2$ if it is a rigid wall.

To make progress with the
problem expressed in terms of averages, we need to relate velocities to $\mathbf{ \bar u}$.
An important feature of the Hartmann layers in this context is that 
the velocity profile in the Hartmann layer is of the form $\mathbf u_\perp=
\mathbf u^cf(z)$, where $f(z)=1-exp(Ha z)$ is the classical Hartmann layer profile
which doesn't depend on the location $(x,y)$.
It follows that the z-average velocity is proportional to the core velocity 
with a constant coefficient: 
\begin{equation}
\mathbf{\bar u}=\mathbf u^c_\perp(1-n \delta^*),
\nonumber
\end{equation}
where $\delta^*$ is the displacement thickness of each Hartmann layer and equal to $Ha^{-1}$. This simple form also implies that the friction $\mathbf{\tau }_{W}$ acts as a linear damping proportional to
the velocity,  with dimensional characteristic time $t_H=(a^2/\nu)(1/Ha)$.
Now neglecting the Reynolds stress of order $Ha^{-1}N^{-1}$ for this particular profile,
(\ref{2d integrate motion equation}) yields the so-called SM82
model in non-dimensional variables:
\begin{equation}
(\partial _{t}+\mathbf{\bar{u}}_{\bot }.\mathbf{\nabla}_\bot)\mathbf{\bar{u}}%
_{\bot }+\mathbf{\nabla }\bar{p}-\frac{N}{Ha^2} \mathbf{\nabla_\bot^2 \bar{u}}_{\bot }
= \frac{N}{Ha}(\mathbf{u}_{0}-n\mathbf{\bar{u}}_{\bot }).
\label{0-ordrer integrate equation}
\end{equation}\\
%

The theoretical precision of this model is 
first order, \textit{i.e.} an error of order $max(1/Ha,1/N)$ is expected on the 
velocity and pressure. In spite of its simplicity, this model is
found to give good results in many well known cases such as parallel layers
(\cite{psm00}) but it fails to describe flows in
which the traditional Hartmann layer is modified  by the presence of inertial effects, such as in
rotating flows for instance. The PSM2000 model described in the next section is
built to overcome this weakness.
\subsection{The PSM2000 model}
\subsubsection{General equations}
In the model developed by \cite{psm00}, a new inertial Hartmann layer profile is derived
from a second order approximation to the Navier-Stokes equations, in the limits
$N \rightarrow \infty$ and $Ha \rightarrow \infty$ (still keeping the ratio 
$Ha/N$ finite). It incorporates inertia as a
 perturbation and is therefore a refinement of the SM82 model. At this order,
  the velocity far from the walls (\textit {i.e.} $z>>Ha^{-1}$ and 
  $1-z>>Ha^{-1}$) is still independent of $z$.  The final 2d model is derived in
  a similar way as SM82, although it involves  more tedious steps. 
The most obvious difference between the PSM2000 and the SM82 model is the appearance of 
cubic terms as well as 
$\partial _t\left(\mathbf{ \bar {u}}_\perp.\mathbf{\nabla }_{\bot }\mathbf{ \bar {u}}_\perp\right)$
terms. They come from the additional terms accounting for inertia in the modified Hartmann 
layer profile. The latter are proportional to $\mathbf{ \bar {u}}_\perp.\mathbf{\nabla }_{\bot }\mathbf{ \bar {u}}_\perp$ and $\partial _t\mathbf{ \bar {u}}_\perp$ so that when this profile is used to evaluate the
 $\overline{\mathbf{u}^\prime.\mathbf{\nabla }\mathbf{ u}^\prime}$ in (\ref{2d integrate motion equation}),
this yields the final form of the PSM2000 equations
\footnote{In fact, the pressure, velocity and time appearing in these 
  equations differ from the averaged quantities by a  constant factor of the form $1+O(Ha ^{-1})$. This small 
  discrepancy is however not relevant here, and is neglected for 
simplicity  throughout the rest  of the paper, as it is not associated to any 
new physical effect.}:
\begin{equation}
\mathbf{\nabla }_{\bot }.\mathbf{\bar {u}}_\perp=0  \label{div v PSM2000}
\end{equation}
\begin{equation}
(\partial _{t}+\mathbf{\bar{u}}_{\bot }.\mathbf{\nabla}_\bot)\mathbf{\bar{u}}_\perp
+\mathbf{\nabla }_{\bot }\bar{p}
-\frac{N}{Ha^2} \mathbf{\nabla }_{\bot}^2\mathbf{ \bar {u}}_\perp
=\frac{N}{Ha}\left(\mathbf{ u}_{0}-n \mathbf {\bar {u}}_\perp\right)  
+\dfrac{n}{HaN}\left( \dfrac{7}{36}\mathcal{D}_{\mathbf {\bar {u}}_\perp}
+\dfrac{1}{8}\partial _t\right) 
\mathbf{ \bar {u}}_\perp.\mathbf{\nabla }_{\bot }\mathbf{ \bar {u}}_\perp 
\label{PSM2000}
\end{equation}
where the operator $\mathcal{D}_{\mathbf{\bar u}_\perp}$ is defined as:
\begin{equation}
\mathcal{D}_{\mathbf{\bar u}_\perp}:\mathbf{F}\longmapsto \mathcal{D}_{\mathbf{\bar u}_\perp}
\mathbf F = \left(\mathbf{\bar u_\perp.\nabla }_{\bot }\right) \mathbf F+\left( 
\mathbf{F.\nabla }_{\bot }\right) \mathbf{\bar u}_\perp.\\
\end{equation}
Out of the two new terms which appear, compared to SM82,
 we are mainly interested in the one with the operator $\mathcal D_{\mathbf{\bar u}_\perp}$, 
 which accounts for the  effects of classical Ekman pumping when a 
 vortex stands over a boundary layer. The advantageous feature of  PSM2000, is that the 
 effects are described  locally, which allows us to determine their influence on any 
 vorticity field. Most of the new results 
 presented in this paper come from the study of this term.\\
 
 The model is more precise than SM82, in the sense that velocity and pressure should
 be evaluated with an error of order $max(1/(HaN),1/N^2, 1/Ha^2)$. In the 
 practical cases studied thereafter, $N$ is in fact smaller than $Ha$ so that 
 the corrections to the velocity involving $1/N$ are more important than 
 those involving $1/Ha$. It can be shown that the terms involving $1/Ha$ merely 
 improve the precision of the model but don't  account for any new phenomenon, 
 as opposed to the $1/N$ terms which carry the effects of the local 3d 
 recirculations (\cite{psm00}).
An analytical model for Hartmann-Bodew\"adt layers can be derived from the present model (for the basic theory of 
Bodew\"adt layers, see \cite{green69}). Comparison 
of the latter with fully non-linear simulations in the axisymmetric case has
shown that (\ref{PSM2000}) is satisfactorily valid if the value of the interaction parameter $N$
remains at least of the order of unity (\cite{dav01}). \\

It should be noticed that 
one of the main advantages of the SM82 and PSM2000 models is that both rely on
asymptotic expansions performed on the Navier-Stokes equation without any 
kind of empirical parameter, which allows us to quantify their precision using non
 dimensional numbers $N^{-1}$ and $Ha^{-1}$.\\
\subsubsection{Effect on the vorticity field}
\label{sec:vortex}
We shall now characterise the PSM2000 model by showing how the local recirculations it accounts for 
affect the vorticity field. The first step consists in deriving  the equation
satisfied by the average vorticity 
$\omega \mathbf e_z=\nabla \times  \mathbf {\bar u}_\perp$ from the 2d model (\ref{PSM2000}). This equation is
obtained by taking the curl of (\ref{PSM2000}), and using the identity
$\mathcal{D}_{\mathbf{\bar u}_\perp}\mathbf{F=\nabla }_{\bot }\left( \mathbf{\bar u_\perp.F}\right)
-\mathbf{\bar u_\perp\times \nabla }_{\bot }\times \mathbf{F-F\times \nabla }_{\bot
}\times \mathbf{\bar u}_\perp$,
as well as  $\mathbf{\nabla }_{\bot }.\omega \mathbf{e}_{z}=0$ :
\begin{multline}
(\partial_t +\mathbf {\bar u}_\perp.\nabla)\omega -
\frac{N}{Ha^2} \nabla_\bot^2\omega = 
-\frac{N}{Ha}\left( \omega _{0}-n\alpha \omega \right) \mathbf{+}
\label{vorticity PSM2000} \\
\tfrac{7}{36}\tfrac{n}{HaN}\left( \left[ \mathbf{\bar u}_\perp.%
\mathbf{\nabla }_{\bot }\mathbf{\bar u}_\perp\right] \mathbf{.\nabla }_{\bot }\omega
+\left( \mathbf{\bar u}_\perp.\mathbf{\nabla }_{\bot }\right) \left(\mathbf{\bar u}_\perp.\nabla_\bot \omega\right)
+\omega \mathbf{\nabla }_{\bot }.\left[ \mathbf{\bar u}_\perp.\mathbf{\nabla }_{\bot }%
\mathbf{\bar u}_\perp\right] \right) 
+\tfrac{1}{8}\tfrac{n}{Ha N}\partial_t \left(\mathbf{\bar u}_\perp.\mathbf{\nabla }_{\bot }\omega\right)
\end{multline}
The additional terms are direct consequences of the secondary flows : as the
non-linear terms  in the expression of the velocity profile in the inertial 
Hartmann layers are proportional to $\mathbf{\bar u}_\perp.\mathbf{\nabla }_{\bot }\mathbf{\bar u}_\perp$ and $\partial _t\mathbf{\bar u}_\perp$, the $%
\left[ \mathbf{\bar u}_\perp.\mathbf{\nabla }_{\bot }\mathbf{\bar u}_\perp\right] \mathbf{.\nabla}
_{\bot }\omega $ and $\partial _t\mathbf{\bar u}_\perp.\mathbf{\nabla }
_{\bot }\omega $ terms represent the amount of vorticity conveyed to the
point $\left( x,y\right) $ from its neighbourhood by secondary flows,
 while the $\left( \mathbf{\bar u}_\perp.\mathbf{\nabla }_{\bot }\right) 
\left(\mathbf{\bar u}_\perp.\bnabla_\bot \omega\right)$ terms represent the transport of vorticity due to these recirculations being carried by the main flow. 
The inertial model of the Hartmann layer also predicts a
vertical velocity proportional to $\mathbf{\nabla }_{\bot }.\left[ \mathbf{\bar u}_\perp%
.\mathbf{\nabla }_{\bot }\mathbf{\bar u}_\perp\right]$ at the edge of the layer. The $%
\omega \mathbf{\nabla }_{\bot }.\left[ \mathbf{\bar u}_\perp.\mathbf{\nabla }_{\bot }%
\mathbf{\bar u}_\perp\right] $ expression is a source term related to the vorticity created
in the core by this phenomenon.\\

The next step is to seek the effects of the non linear terms of (\ref%
{vorticity PSM2000}) on a vortex spot sketched as a local extremum of
vorticity. We assume that the vorticity field exhibits a local extremum and that this 
extremum is conveyed by a background flow $V\mathbf{e}_{x}$. The extremum is thus located
at the point $\left( x_{0}+Vt,y_{0}\right) $ \textit{i.e. }: 
%

\begin{equation}
\partial _{x}\omega \left( x_{0}+Vt,y_{0}\right) =\partial _{y}\omega \left(
x_{0}+Vt,y_{0}\right) =0  \label{cond on dw extremum} \\
\end{equation}

In addition, the background flow $V\mathbf{e}_{x}$ is considered constant 
and large in front of the local velocity variations:
%
\begin{subequations}
\begin{gather}
\mathbf{\bar u}_\perp=V\mathbf{e}_{x}+\mathbf{v}^{\prime }\left( x,y,t\right) \\
\partial _{x}V\left( x,y,t\right) =\partial _{y}V\left( x,y,t\right)=\partial _tV\left( x,y,t\right)=0
\label{cond on V} \\
\left\| \mathbf{v}^{\prime }\left( x,y,t\right) \right\| \ll V
\label{cond on v'}
\end{gather}
\end{subequations}
so that the local velocity $\mathbf{v}^{\prime }\left( x,y,t\right)=v_x 
\mathbf{e_x}+v_y\mathbf{e_y} $ satisfies the conservation equation:
\begin{equation}
\mathbf{\nabla }_{\bot }.\mathbf{v}^{\prime}=0
\label{div v local}
\end{equation}
The extremum condition (\ref{cond on dw extremum}) implies that
the transport by secondary flows doesn't act:
\begin{equation}
\left[ \mathbf{\bar u}_\perp.\mathbf{\nabla }_{\bot }\mathbf{\bar u}_\perp\right] 
\mathbf{.\nabla }_{\bot }\omega \left( x_{0}+Vt,y_{0}\right) =0. \\
\end{equation}
Expanding $\omega \mathbf{\nabla }_{\bot }.\left[ \mathbf{%
\bar u}_\perp.\mathbf{\nabla }_{\bot }\mathbf{\bar u}_\perp\right] \left( x_{0}+Vt,y_{0}\right) $ and $\left( \mathbf{\bar u}_\perp.\mathbf{\nabla }_{\bot }\right) \left(\mathbf{\bar u}_\perp.\bnabla_\bot \omega\right) \left( x_{0}+Vt,y_{0}\right) $ in terms of the
derivatives of $\omega$ and $\mathbf{v}^{\prime },$ and using (\ref{cond on
dw extremum}), (\ref{cond on V}) and (\ref{div v local}) yields:
\begin{subequations}
\begin{gather}
\omega \mathbf{\nabla }_{\bot }.\left[ \mathbf{\bar u}_\perp.\mathbf{\nabla }_{\bot }%
\mathbf{\bar u}_\perp\right] \left( x_{0}+Vt,y_{0}\right) =
2V^{2}\partial _{xx}^{2}\omega
+\omega \left[ 2\left( \partial _{x}v_{x}^{\prime }\right) ^{2}+\left(
\partial _{x}v_{y}^{\prime }\right) ^{2}+\left( \partial _{y}v_{x}^{\prime
}\right) ^{2}-\omega ^{2}\right] .
\label{eq:nl1}\\
\left( \mathbf{\bar u}_\perp.\mathbf{\nabla }_{\bot }\right) \left(\mathbf{\bar u}_\perp.\bnabla_\bot \omega\right) \left( x_{0}+Vt,y_{0}\right)=
V^{2}\partial _{xx}^{2}\omega+2V(\partial_x \mathbf v^\prime).\bnabla_\bot \omega+
+V\mathbf v^\prime.\bnabla_\bot\partial_x\omega +\mathbf v^\prime.\bnabla_\bot(\mathbf v^\prime.\bnabla_\bot \omega).
\label{nl at an extremum}
\end{gather}
\end{subequations}%
Using the relation $\partial_t=V\partial_x$ for the advected extremum, the unsteady term can be rewritten as:
\begin{equation}
\partial _t\left(\mathbf{\bar u}_\perp.\mathbf{\nabla }_\bot\omega\right) \left(x_0+Vt,y_0\right)=
V^2\partial^2_{xx}\omega+
V(\partial_x \mathbf v^\prime).\bnabla_\bot\omega +V\mathbf v^\prime.\bnabla_\bot\partial_x\omega.
\label{eq:uns_extremum}
\end{equation}
The condition (\ref{cond on v'}) ensures that the terms proportional to $V^2$ 
in the \textit{%
r.h.s. }of (\ref{eq:nl1}), (\ref{nl at an extremum}) and (\ref{eq:uns_extremum}) are 
arbitrarily larger than the others. Then $\omega \mathbf{\nabla }_{\bot }.%
\left[ \mathbf{\bar u}_\perp.\mathbf{\nabla }_{\bot }\mathbf{\bar u}_\perp\right] \left(
x_{0}+Vt,y_{0}\right) \simeq 2V^{2}\partial _{xx}^{2}\omega $,
$\left( \mathbf{\bar u}_\perp.\mathbf{\nabla }_{\bot }\right) \left(\mathbf{\bar u}_\perp.\bnabla_\bot \omega\right) \left( x_{0}+Vt,y_{0}\right) 
\simeq V^{2}\partial _{xx}^{2}\omega$ and 
$\partial _t\left(\mathbf{\bar u}_\perp.\mathbf{\nabla }_\bot\omega\right) \left(x_0+Vt,y_0\right) \simeq V^{2}\partial _{xx}^{2}\omega$ so that the non
linear term acts on the vorticity as an anisotropic diffusion, in the
direction of the background velocity, with related diffusivity:
\begin{equation}
\eta =\frac{17n}{24}\frac{Ha}{N_V^{2}}\nu 
\label{nl diffusivity}
\end{equation}
where, $N_V$ is the interaction parameter based on the  average flow $V$. This 
confirms that the additional terms are dissipative, as shown by 
\cite{dellar03}. The related diffusivity can also be seen as a turbulent
diffusivity which is determined the by secondary flows. This extends 
the analogy with the usual turbulent Reynolds stresses which are sometimes
interpreted as a turbulent diffusion with related "eddy viscosity".\\

The main result of this section is that any elementary
vortex of the flow is spread by nonlinearities, so that the latter have a
smoothing effect on the whole velocity field. This is to be related to the 
results of the numerical simulations presented in section \ref{simul_matur}.
Also, \cite{dellar03} showed that the non-linear terms in 
(\ref{PSM2000}) 
induce a diffusion along streamlines for small amplitude waves. He also showed that
PSM2000 shares this feature with the model proposed by
\cite{benzi90} for two-dimensional turbulence based on ideas from the 
Anticipated Vorticity Method of \cite{basdevant83}. This model features additional 
non linear terms like PSM2000 and describes 
well some oceanic and atmospheric flows.
This suggests that accounting for the 3d recirculations in oceans and 
atmospheres could lead to accurate 2d models similar to PSM2000.

\section{Numerical setup}
\label{sec:numeric}
\subsection{The numerical model}
We use the finite volume code FLUENT/UNS featuring  a second order upwind spatial 
discretisation. The cases studied are unsteady and the time-scheme is a second 
order implicit pressure-velocity 
formulation. Within each iteration,  equations are solved one after the other (segregated mode)
 using the PISO algorithm proposed by \cite{issa85}. In short, PISO is a 
 predictor-corrector method which substantially reduces the number of
 iterations per time step, especially  in unsteady calculations, by decomposing
  each iteration into one prediction step and several (two here) correction 
  steps: in the prediction step, a first (predicted) velocity field is obtained by 
 solving the momentum equations in which the value of the pressure is taken 
 from the result of the previous iteration (the equations are then implicit for the
velocity but explicit for the pressure). 
 In the next step, a corrected  pressure is obtained by solving an explicit Poisson 
 equation, in  which the velocity is the result from the prediction step.
 A second (corrected) velocity field is solution of the momentum equations 
 in which inertial terms are evaluated using the velocity obtained in the prediction 
 step and in which the pressure is the corrected one. This last step (called 
 correction) is iterated one additional time. Note that this algorithm is in fact
 a modified  version of the one described by \cite{issa85} in which the prediction-correction 
is applied in between time steps rather than in between iterations within the same time step.\\
 	The additional terms in (\ref{PSM2000}) are modelled the following way:
\begin{itemize}
\item[-]The Hartmann friction $ -\mathbf{\bar u}_\perp/t_H$ is expressed implicitly, \textit{i.e.}
 as $-\mathbf{\bar u}^{(n+1)}_\perp/t_H$ at current time $t^{(n+1)}$, where 
$\mathbf{\bar u}^{(n+1)}_\perp$  is the velocity variable at the current time step, on
which the PISO iterations are performed.
\item[-]The $\mathbf{\bar u}_\perp\bnabla_\bot.\mathbf{\bar u}_\perp$ terms appearing in the PSM2000
model additional terms and their gradients are treated implicitly in time and updated
at the end of each iteration within the time steps, using the latest values of the velocity 
obtained from the resolution of the pressure-velocity equations by the PISO algorithm.
These terms are therefore not modified during the PISO iterations.
\item[-]The additional time derivative is second order implicit, \textit{i.e.}
expressed at time $t^{(n+1)}$ as $[\partial_t (\mathbf {\bar u}_\perp.\bnabla_\bot) \mathbf {\bar u}_\perp]_{t^{(n+1)}}=
\frac{1}{2\Delta t}\left(3[(\mathbf {\bar u}_\perp.\bnabla_\bot) \mathbf {\bar u}_\perp]_{t^{(n+1)}}
-4[(\mathbf {\bar u}_\perp.\bnabla_\bot) \mathbf {\bar u}_\perp]_{t^{(n)}}
+(\mathbf {\bar u}_\perp.\bnabla_\bot) \mathbf {\bar u}_\perp]_{t^{(n-1)}}\right)$, where the superscripts $(n)$ and $(n-1)$ refer to the variables taken from the 
two previous time steps.
\end{itemize} 
\begin{figure}
\centering
\includegraphics[scale=0.85]{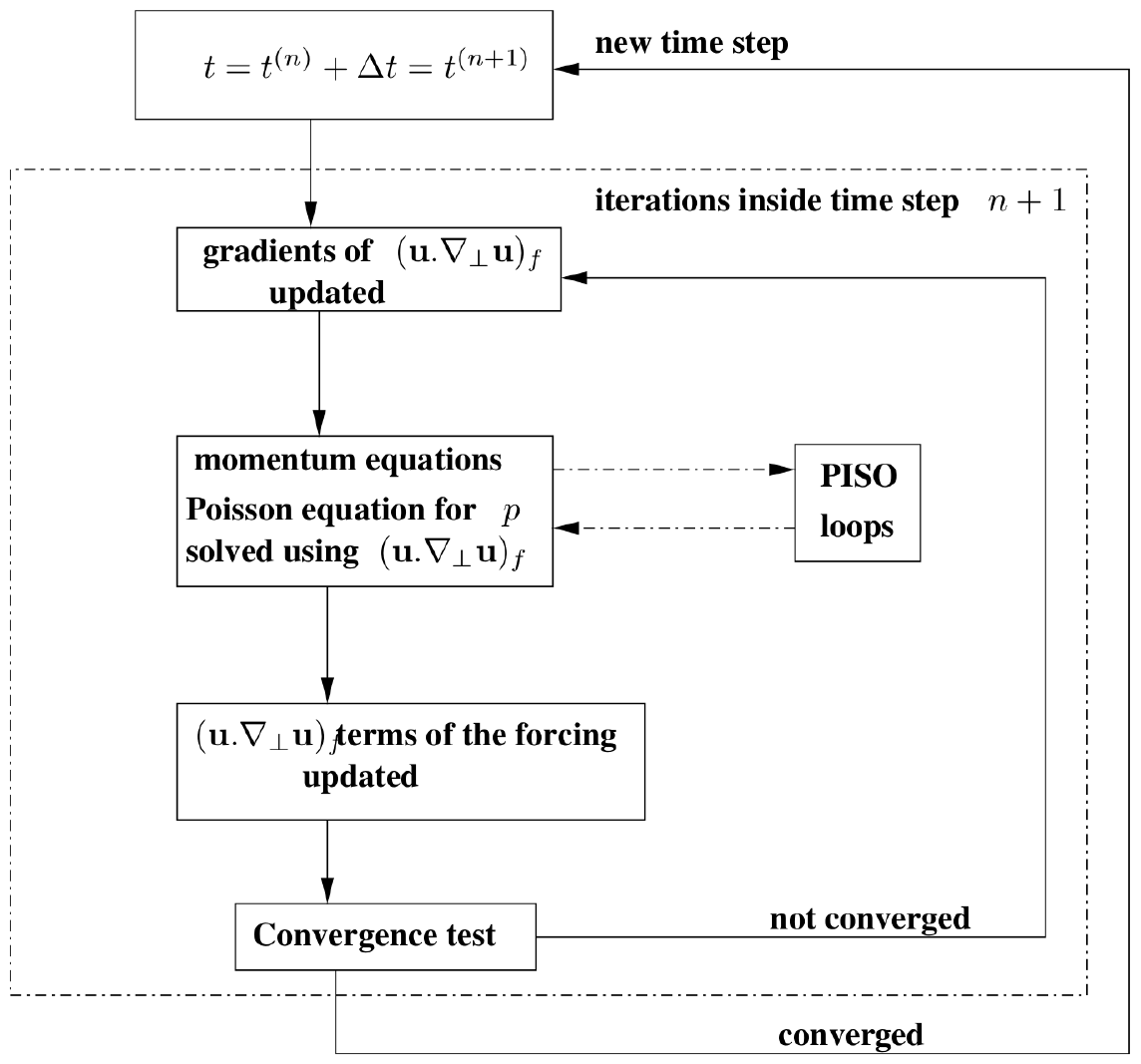}
\caption{Algorithm used to solve the equation (\ref{PSM2000}) numerically. 
$(\mathbf u.\bnabla_\bot\mathbf u)_f$ represents the inertial terms appearing in the 
additional terms of PSM2000.}
\label{fig:algo}
\end{figure}

A summary of the algorithm is sketched in figure \ref{fig:algo}.
\subsection{Tests on the numerical model}
\begin{figure}
\centering
\includegraphics[scale=0.5]{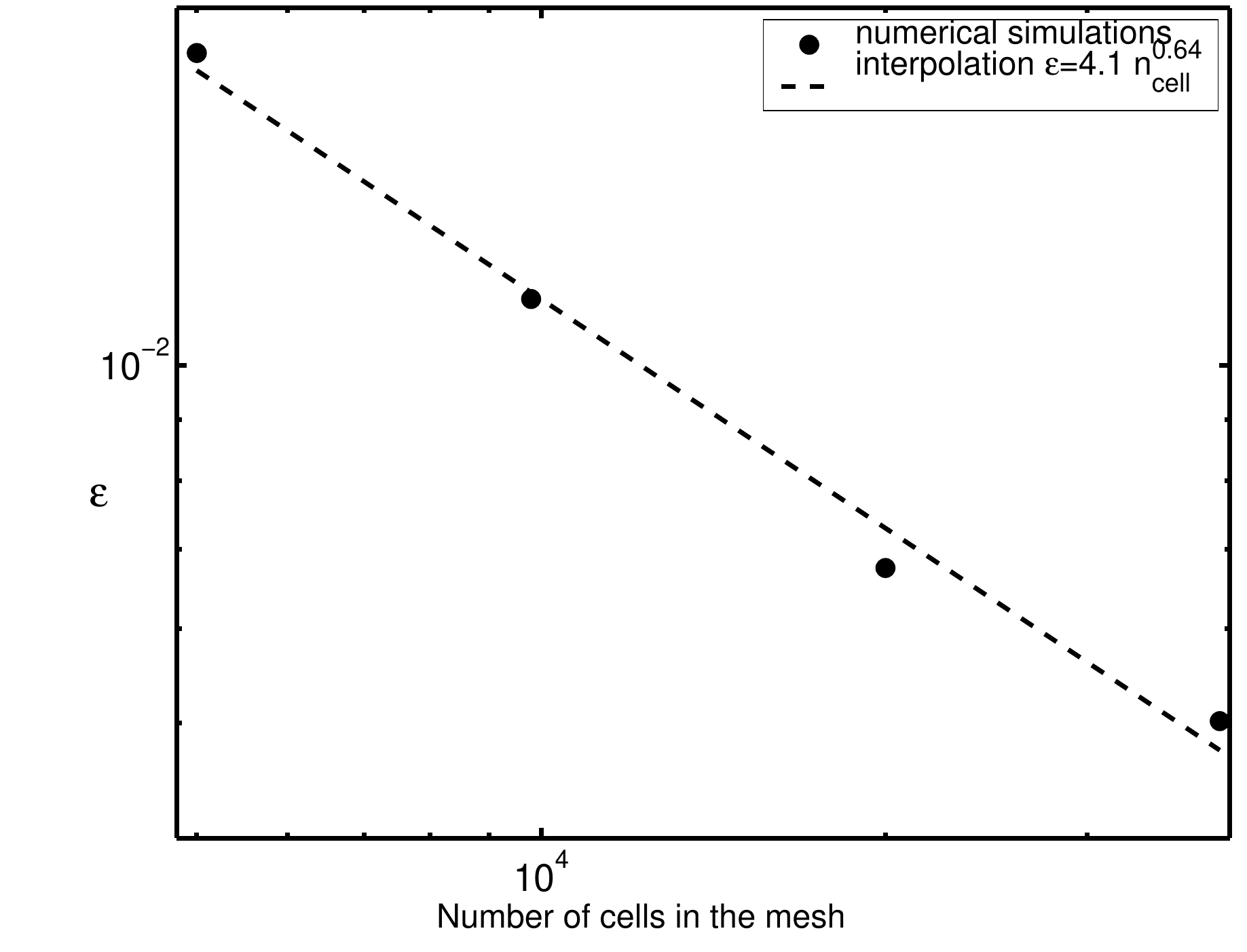}
\caption{Time average of the $\mathcal L_2$ norm of the relative error 
$\epsilon=\|U_{numeric}-U_{ref}\|_2/\|U_{ref}\|_2$ 
in numerical simulations, compared
to the reference analytical solution (\ref{eq:ref_anal}) versus number of cells in the mesh.}
\label{fig:convergence}
\end{figure}

We shall now investigate the ability of the numerical system to solve equations
(\ref{PSM2000}). To this end, we perform a convergence 
study under grid refinement toward an analytical solution. As these equations 
 are both new and complex, no exact analytical solution has been 
exhibited up 
to now. We therefore follow the procedure  recommended
by \cite{roache97} which consists in specifying an analytical velocity field and
adjusting the forcing term (here $\mathbf u_0$ in (\ref{PSM2000})) so that the 
specified field is solution of the equations. We choose the case of a flow 
confined between two co-rotating vertical cylinders (respective radius $r_{int}$
 and $r_{ext}$) and two horizontal plates 
(at $z=0$ and $z=1$), plunged in a vertical uniform magnetic field. 
Equations (\ref{PSM2000}) then apply on the 2d annulus $r_{int}<r<r_{ext}$. 
The parameters $r_{int}$,
$r_{ext}$, $Ha$ and $N$ can be set for the solution to exhibit a 
significant Ekman pumping, which is the very kind of phenomenon the PSM2000 model  
is supposed to account for. The reference solution consists in an azimuthal
 wave superimposed on a $1/r^2$ axisymmetric radial profile. Numerical constants 
are adjusted so that the wave amplitude is $10\%$ of the azimuthal velocity at 
the inner cylinder, and so that the velocity is tangent to the walls located at
$r=r_{int}$ and $r=r_{ext}$:
\begin{eqnarray}
\frac{v_\theta(r)}{v_\theta(r_{int})}=\frac{r^2_{int}}{r^2}+
0.407(r-r_{int})(r-r_{ext})(r-\frac{r_{int}+r_{ext}}{2})cos(7\theta+3.5t)  \nonumber \\
\frac{v_r(r)}{v_\theta(r_{int})}=\frac{2.4929}{r}(r^2_{int}r^2_{ext}
-2rr_{int}r_{ext}(r_{int}+r_{ext})+r^2(r^2_{int}+r^2_{ext}+4r_{int}r_{ext}- \nonumber \\
2r^3(r_{int}+r_{ext})+r^4)sin(7\theta+3.5t) 
\label{eq:ref_anal}
\end{eqnarray}
The initial conditions at $t=0$ and the Dirichlet boundary conditions at the walls for the 
velocity are chosen to match (\ref{eq:ref_anal}).
These conditions avoid the occurrence of a boundary layer along the cylinders 
for which no analytical solution would be known.
Setting $Ha=111$, $N=12$ ($N$  is built using $v_\theta(r_{int})$  and 
$r_{ext}/r_{int}=10$ ensures that the region $r_{ext}-r<<r_{ext}$ is dominated by 
viscosity while non-linear terms dominate the dynamics near the inner cylinder
 $r_{int}-r<<r_{int}$). The convergence tests are performed on a structured mesh
 with twice as many azimuthal nodes as along one radius. The time steps are adjusted to satisfy the 
Courant-Friedrich-Lewy condition for the maximal azimuthal phase velocity of the imposed wave 
(\textit{resp.} 0.018 s, 0.0128 s, 0.009 s, 0.007 s for cases with \textit{resp.} 50, 70, 100 
and 140 radial modes). 
Each calculation 
runs over a full time-period of the imposed solution.  Figure \ref{fig:convergence}
 shows that the $\mathcal L_2$ norm of the relative error over the domain 
 decreases approximately 
  as $n_{cell}^{-0.64}$ where $n_{cell}$ is the number of elements in the mesh. This 
  confirms the reliability of the numerical system. It is however important to notice that 
the convergence is not of second order spatial accuracy, although all quantities are being 
discretised at this order. The reason for this precision loss is that the 
$\mathcal{D}_{\mathbf {\bar {u}}_\perp}\mathbf{ \bar {u}}_\perp.\mathbf{\nabla }_{\bot }\mathbf{ \bar {u}}_\perp$
 terms appearing
in (\ref{PSM2000})  are 
calculated by taking 
the gradient of the  $\mathbf{\bar u}_\perp.\bnabla_\bot \mathbf{\bar u}_\perp$ variables. Although 
these 
variables are known to second order precision, the resulting gradients are not. 
The achieved accuracy is
however sufficient for our purpose, which is to model physical experiments 
rather than
to build a refined numerical model. Such a refined numerical work based on the PSM2000 model can be 
found in \cite{dellar03}.
\section{Simulation of the free decay of isolated vortices generated by a
single-electrode}
\label{sec:sommeria}
\subsection{Experimental device of reference}
\begin{figure}
\centering
\includegraphics[width=0.65\textwidth]{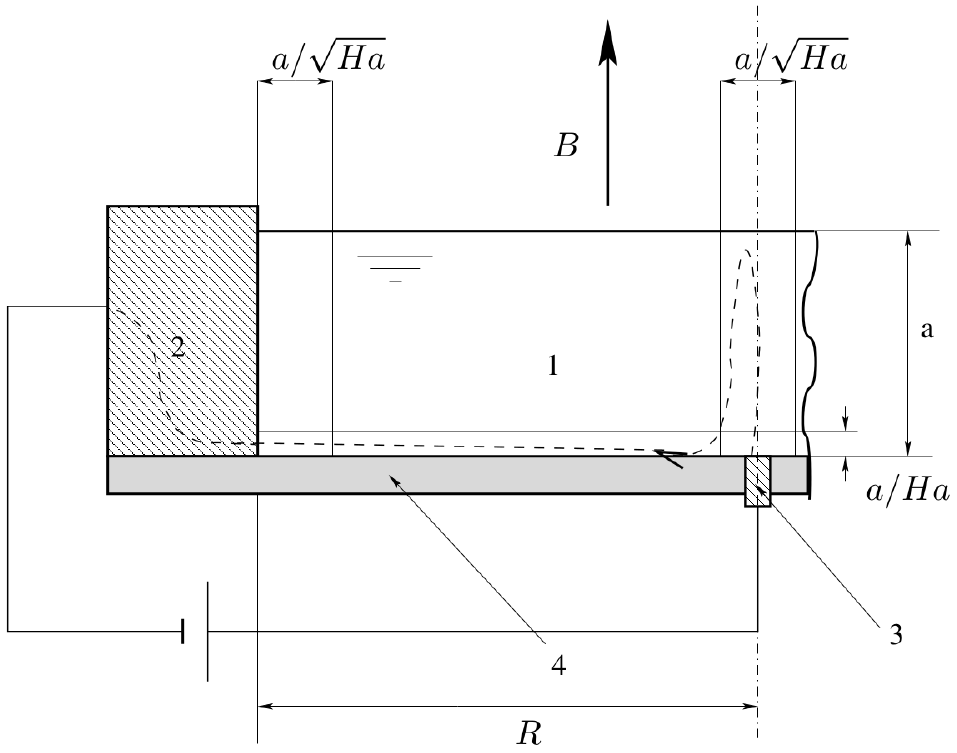}
\caption{Experimental device of Sommeria's vortex study: cross section of the 
circular tank with a schematic representation of the current supply. 
A typical current 
streamline passing through the Hartmann layer is also 
represented  (dashed). 1: mercury, 2: electrically conducting side wall, 3: electrode for 
current injection, 4: electrically insulating bottom wall.}
\label{expsommeria}
\end{figure}

In the next two sections, we shall use the numerical implementation of 
both PSM2000 and SM82 described in the previous section to recover the 
results of two MHD experiments, which couldn't be modelled by classical 
theories. We first perform the simulations on \cite{sommeria88} 's electrically 
driven vortices using PSM2000 only. The experimental setup consists in a 
cylindrical tank (diameter $2R=120$ mm) filled with mercury (depth $a=19.2$ mm)
 with an insulating bottom plate, an upper free surface ($n=1$) and an electrically 
 conducting circular wall at $r=R$ (see figure \ref{expsommeria}). Electric 
 current is injected into the mercury via a small electrode (diameter $2r_e=2.5$ mm)
  located in the bottom plate. The injected current $j_W$ can be approximated 
 as a Dirac-delta function centred at the edge of the electrode $r=r_e$, with
 integral equal to the total injected current $I$: 
 $j_W=I/(2\pi r_e)\delta(r-r_e)$.
 The corresponding forcing is azimuthal and given from the solution of 
 (\ref{current}) which yields:
\begin{equation}
\forall  r>r_e, \mathbf{u}_{0}=-\dfrac{B}{\rho a}\dfrac{I}{2\pi r}%
t_{H}\mathbf{e}_{\theta }.  \label{sommeria forcing}
\end{equation}
The forcing is applied until a steady regime is reached. This flow is quite
stable and remains laminar. At the end of the run, the forcing is switched off and the flow decays by Hartmann friction. The experimental parameters are 
summarised in the table below with the corresponding non-dimensional 
parameters and numerical time-steps.
(We give here the values of $N_c=N/\sqrt{Ha}$, which is scaled 
on the vortex core thickness of order $a Ha^{-1/2}$, as \cite{sommeria88} 
noticed that 
it is the relevant parameter that governs the recirculating effects in the vortex):
\begin{center}
\begin{tabular}{|p{4cm}|p{2cm}p{2cm}p{2cm}p{2cm}|}
\hline
$B/$T		& 0.0575 & 0.115 & 0.23 & 0.48\\
\hline
$Ha$ 		& 28.41 & 56.82 & 113.6 & 237.2\\
$t_H$/s 	&110.9 & 55.45 & 27.72 & 13.28\\
$N_c$ ($I=50$ mA) 	& 0.017 & 0.034 & 0.068 & 0.14\\
$N_c$ ($I=12.5$ mA)	& & & & 0.569\\
$N$ ($I=50$ mA) 	& 0.091 & 0.256 & 0.724 & 2.16\\
$N$ ($I=50$ mA) 	& & & & 8.76\\
time step /s ($I=50$ mA) &0.043&0.065&0.076&0.080\\
time step /s ($I=12.5$ mA)& & & &0.086\\
\hline
\end{tabular}
\end{center}
\subsection{Mesh and boundary conditions}
The mesh is made of quadrilateral elements, unstructured for $r<1.64$ mm and structured 
for $1.64$ mm$<r<R$. The radial resolution is 105 points, 25 of which are devoted to
the boundary layer located at $r=R$. These points are spread in the layer according to 
 a geometric sequence of ratio 1.3 starting at $r=R$ with an initial interval
of $5.10^{-6}$ mm.  The azimuthal resolution is of 150 points. The time step is 
chosen so that the related cutoff frequency matches the spatial cutoff frequency
for the maximal flow velocity (Courant-Friedrich-Lewy condition). 
Values are given in the table above. The usual no-slip condition at the wall $r=R$ is applied.
\subsection{Free decay}
\label{sec:decay}
\begin{figure}
\begin{center}
\includegraphics[scale=0.7]{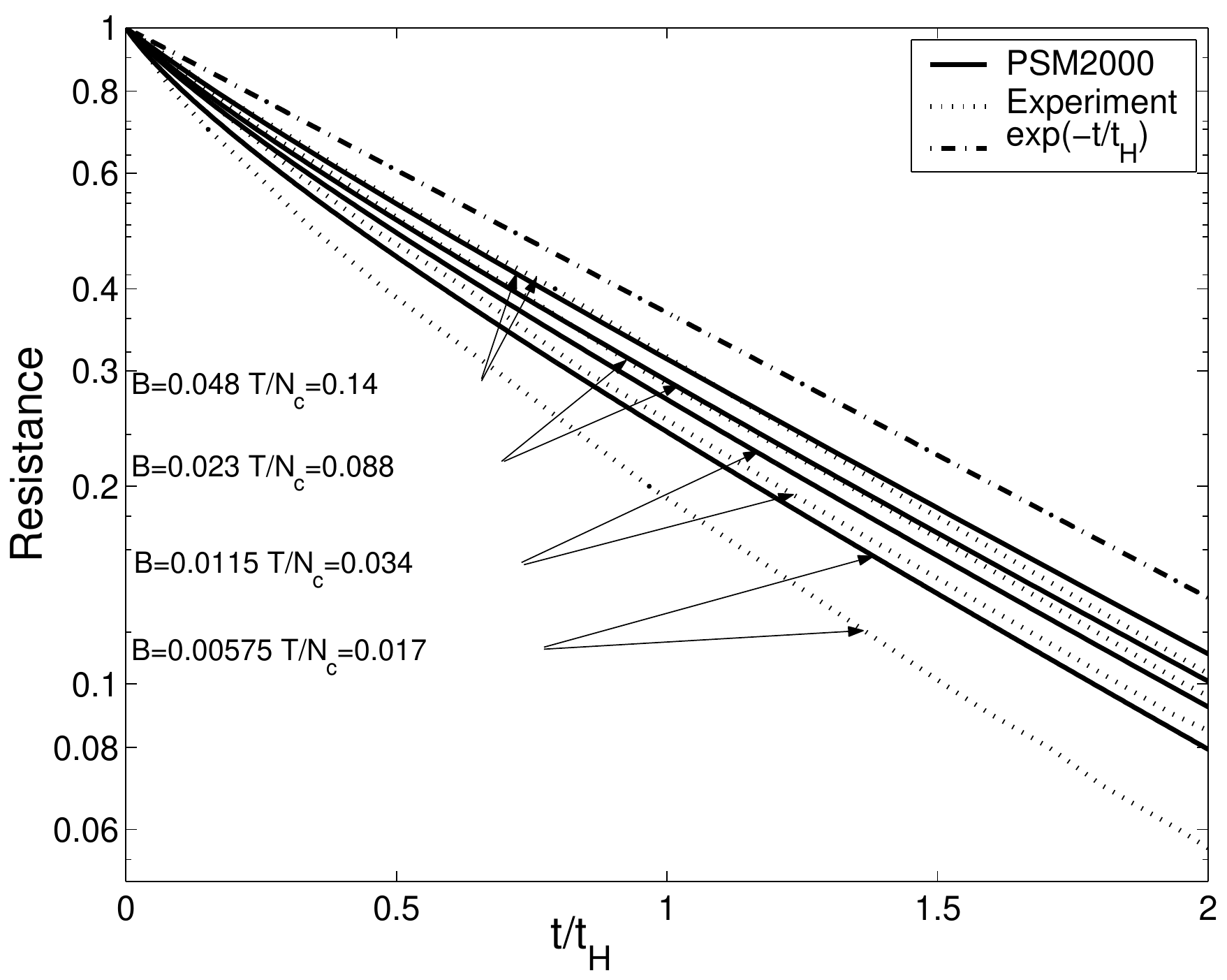}
\caption{Time-decay of the resistance between the central electrode and the side
 wall for several magnetic fields (injected current before decay is $I=0.05$ A 
 unless otherwise specified). 
 Resistances are normalised by their value at equilibrium, 
at the moment the forcing is switched off ($t=0$ on the graphs). The exponential
decay is the one predicted by the SM82 model.}
\label{tourbint}
\end{center}
\end{figure}

%
%
Figure \ref{tourbint} shows the decay of electric resistance between the
central electrode and the conductive side wall. This quantity is derived in \cite{sommeria88} from
the velocity field as $\mathcal R=(\tilde\psi_{wall}-\tilde\psi_{electrode})/I=-1/(IB)\int_b^R u_\theta dr$, 
(where $\tilde\psi$ is the dimensional electric potential) using the fact that there 
is no current outside the Hartmann layer because the flow is two-dimensional. 
The numerical simulations from the model show
that $\mathcal R$ decays strongly at early times, and the decay rate then 
stabilises 
around $(0.9t_H)^{-1}$. This agrees very well with the experiment. 
Also, the small discrepancy between PSM2000 and the experiment increases with 
$N_c^{-1}$. 
This is precisely what one should expect as PSM2000 is derived from asymptotic 
expansions on $Ha^{-1}$ and $N^{-1}$. This tends to confirm that these 
non-dimensional 
parameters provide a good measure of the precision of the model.\\

\begin{figure}
\centering
\includegraphics[scale=0.8]{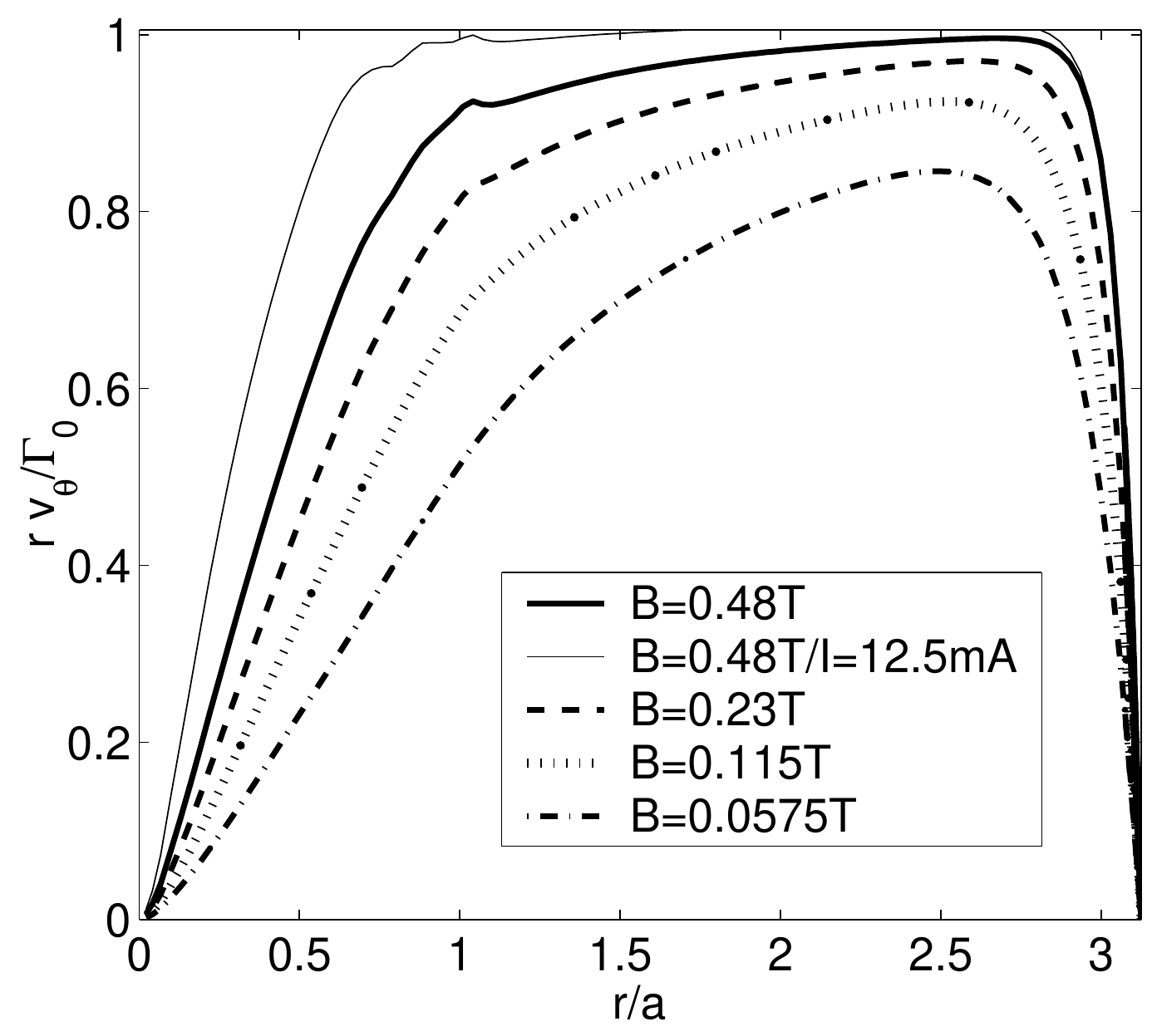}
\includegraphics[scale=0.65]{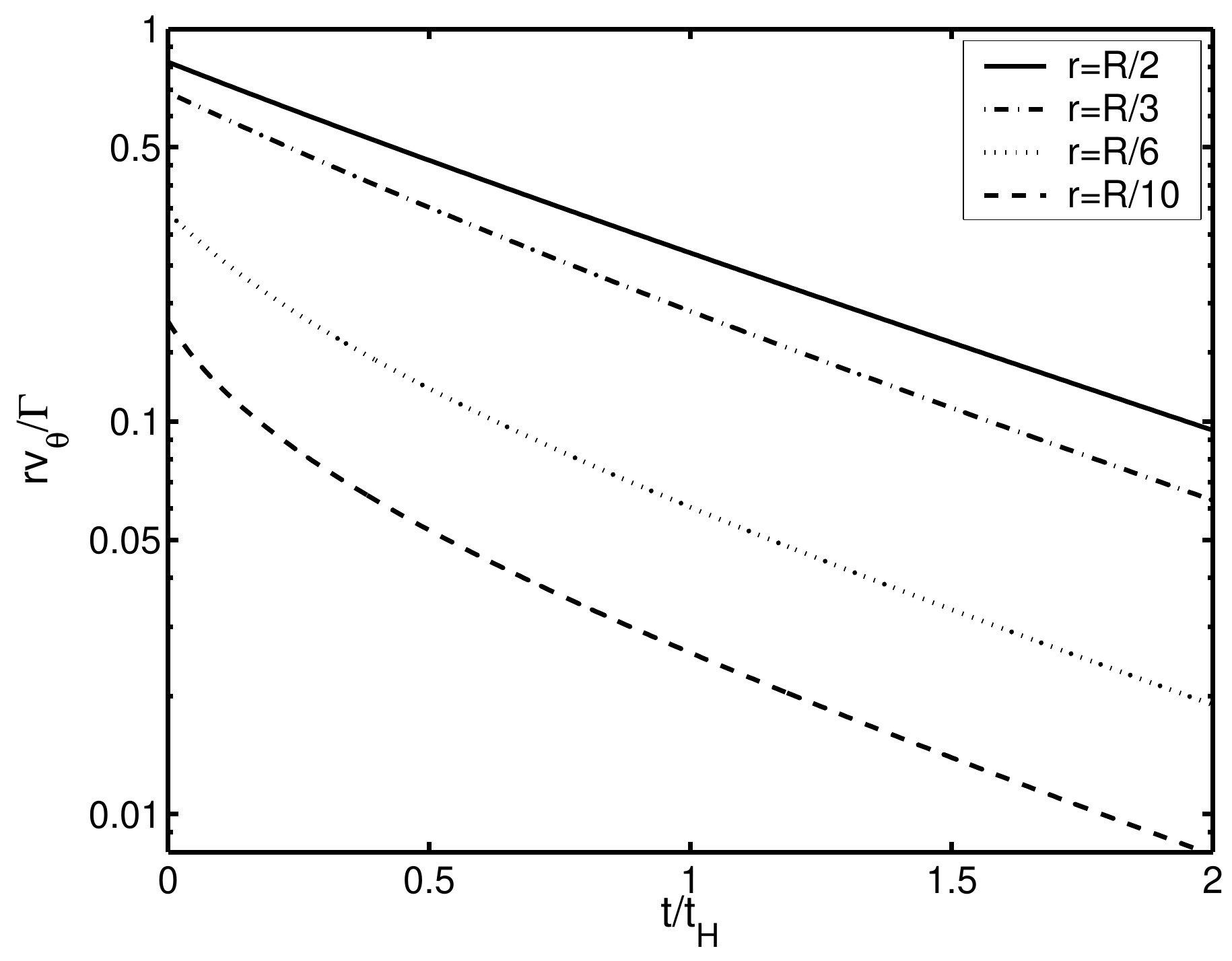}
\caption{Angular momentum normalised by $\Gamma_0=I/(2\pi(\rho\sigma\nu)^{1/2})$.
The radial profiles 
of angular momentum (top) are obtained numerically and show that the main effect of the recirculations is to
broaden the vortex core. Injected current before decay is $I=0.05$ A
 unless otherwise specified.
Time-decay of azimuthal velocity for different radial positions (bottom), just 
after the forcing is switched off (referred to as $t=0$ here), obtained from the numerical simulations 
of the PSM2000 model at B=0.23 T. Secondary flows are stronger at the centre, where the velocity decays faster. 
}
\label{vtourb}
\end{figure}

Physically, the
strong damping at early times - weaker for weak currents and strong fields -
is explained by the presence of Ekman recirculations. Indeed, Ekman pumping 
induces a centrifugal flow in the core flow as well as a
centripetal flow in the Hartmann layers. The mass conservation requires
that the vertically integrated  mass fluxes related to these two radial flows 
be the same. As the velocity is smaller in the
Hartmann layer, the net effect of Ekman pumping is a centrifugal
transport of angular momentum. This has two consequences: The first one is that
the wall 
side boundary layer is
squeezed by this transport so that the wall friction is increased. The 
recirculations are important when 
the vortex still rotates fast, so that
angular momentum is conveyed toward the side layer, which increases
dissipation and enhances the damping. This phenomenon is however not very 
strong in the present case since the velocities near the lateral wall are
rather small, as opposed to the MATUR case described in section \ref{simul_matur}.\\
When the flow has been significantly
damped, the Ekman recirculation disappears and the wall side layer goes back to
its typical $Ha^{-1/2}$ thickness so that the associated dissipation becomes 
small compared to the Hartmann damping. The decay rate of the velocity then matches
approximately the $t_{H}^{-1}$ value predicted by the linear theory. 
The second consequence is 
that azimuthal velocities initially decrease much faster for points
which are closer to the centre as shown in figure \ref{vtourb}. The 
reason is that the recirculations arise from centripetal jets in the
Hartmann layer, which are therefore stronger at the centre of the tank. This 
also explains that recirculations
tend to noticeably "broaden" the vortex core, as measured  by \cite{sommeria88} and 
confirmed theoretically by \cite{psm00}.
\section{Numerical Simulations for the MATUR\ experimental setup}
\label{simul_matur}
\subsection{Experimental device of reference}
\begin{figure}
\begin{center}
\includegraphics[width=0.7\textwidth]{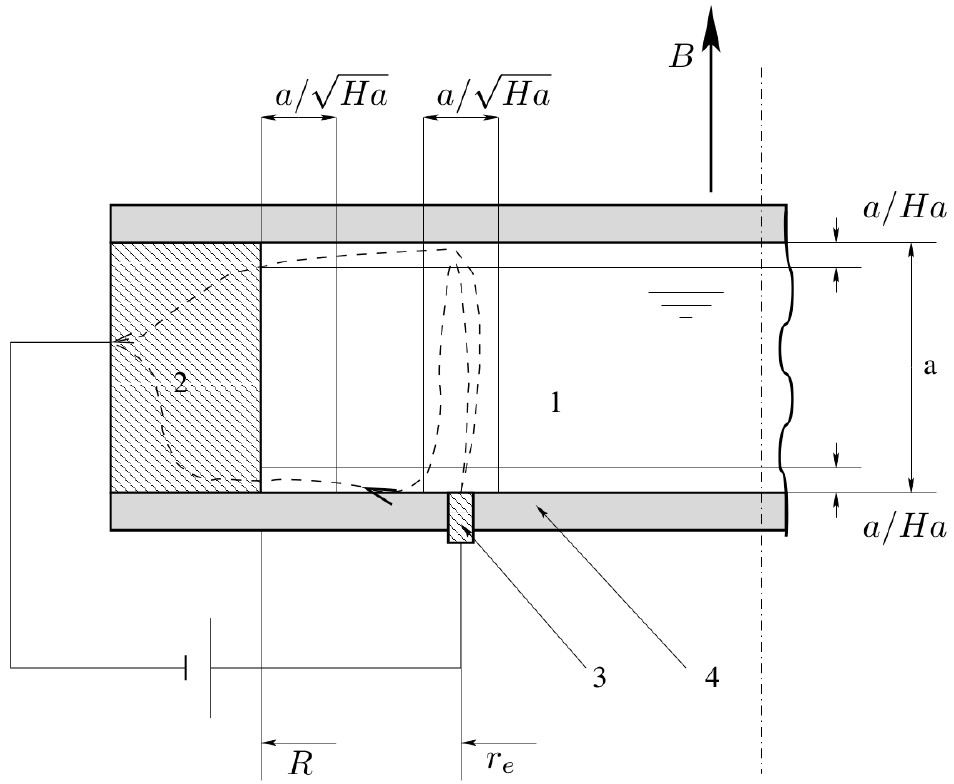}
\caption{Radial section of the MATUR
experimental setup. Some typical current
streamlines passing through the Hartmann layers are also
represented  (dashed). 1: mercury, 2: electrically conducting side wall, 3: current
injection electrode, 4: electrically insulating bottom wall.}
\label{Matur scheme}
\end{center}
\end{figure}

We now come to the main part of this work, where PSM2000 and SM82 are both 
compared and used to recover and explain the results obtained by
\cite{albouss99} with the MATUR (MAgnetic TURbulence) experimental setup 
developed in Grenoble. MATUR is a cylindric container (diameter $2R=0.22$ m)
with an electrically
insulating bottom and conducting vertical walls (figure \ref{Matur scheme}).
Electric current is injected at the bottom through a large number of
point-electrodes regularly spread along a circle the centre of which is on the
axis of the cylinder. It is filled with mercury ($a=1$ cm depth) and the whole
device is  placed in a steady uniform vertical magnetic field. 
The injected current leaves
the fluid through the vertical wall inducing radial electric current lines
and gives rise to and azimuthal force on the fluid included in the annulus
between the electrode circle and the outer wall.\\

The forcing is similar to the case of section \ref{sec:sommeria} but the radius
where the current is injected $r_e=0.093$ mm is much larger so that a free shear
layer is produced with a vorticity sheet at $r=r_e$. Instability is associated
 with this vorticity extremum. By contrast, in the case of section 
 \ref{sec:sommeria}, the vorticity extremum was at the centre of the tank,
 leading to a stable flow.\\

%
%
The annulus of fluid $r \in [r_e, R]$ rotates and gives rise to a concave
 parallel wall side layer along the outer wall ($r=R$) and a free parallel shear
 layer at $r=r_e$. The upper surface is rigid so that two Hartmann
layers (at the top and the bottom) are present ($n=2$ ).\\
The field is $B=0.17$ T (\textit{i.e.} $Ha=45.14$) and the fluid is at rest 
at the initial state $t=0$.
Numerical simulations are performed for a total injected current $I$ in the
range $[3$ A$,30$ A$]$.
An approximate azimuthal velocity $U_{sm82}=I/(2\pi R \sqrt{\sigma \rho \nu})$ 
and associated global angular momentum per unit of height  $L_{sm82}=U_{sm82}\pi R(R^2-re^2)$ 
can be derived from the theory from \cite{sm82} (see \cite{psm00}), the order of magnitude of 
which remains
valid within the framework of PSM2000. The relevant interaction parameter is scaled 
on the horizontal length $N_{2d}=\sigma B^2 R/(\rho U_{sm82})$.
The horizontal velocity $U_{sm82}$ is used for convenience, but this is an overestimate, so that the physical interaction parameter should be somewhat higher 
than $N_{2d}$.\\
\begin{center}
\begin{tabular}{|p{3cm}|p{1cm} @{}p{1cm}@{}p{1cm}@{}p{1cm}|}
\hline
$I$ /A		&3	&10	&20	&30	\\ 
$U_{sm82}$ /m/s	&0.054	&0.18	&0.36	&0.54	\\
$N_{2d}$		&4.6	&1.4	&0.67	&0.47	\\
time step  /s &0.025&0.01&0.01&0.007\\
\hline
\end{tabular}
\label{tab:matur}
\end{center}
A more
comprehensive description of the experimental device and results can be found
in \cite{albouss99}.
\subsection{Numerical setup}
As the geometry is similar to that of Sommeria's experiments described 
in section 
\ref{sec:sommeria}, we use the same mesh and the same boundary conditions at the wall
located at $r=R$. This mesh ensures that the 
wall side layer located at $r=R$ is always described by at least 12 points. 
In order to reduce the CPU time, the free shear layer located at $r=r_e$
 is not finely meshed. Indeed, the latter is thin in laminar regime
  (thickness $a Ha^{1/2}$) which only happens in the first few seconds of
   each case (out of more than one minute duration of the real experiment). The layer 
   then quickly destabilises and is 
 replaced by large vortices with relatively smooth velocity gradients which
 do not require mesh refinement. This simplification might 
 make the modelled layer slightly more unstable than the real one 
 but does not significantly affect the quasi-steady state we are mostly 
 interested in. As in section \ref{sec:sommeria}, The time 
step is chosen to satisfy the Courant-Friedrichs-Lewy condition, so that the
temporal cutoff frequency matches the spatial cutoff 
frequency (see table \ref{tab:matur}).\\
All time-averaged values are calculated in the steady regime reached after a
time of $3t_H$. The statistics are then performed over a period of $t_H$.\\
\subsection{Overview of the simulated flow}
\begin{figure}
\centering
\includegraphics[width=\textwidth]{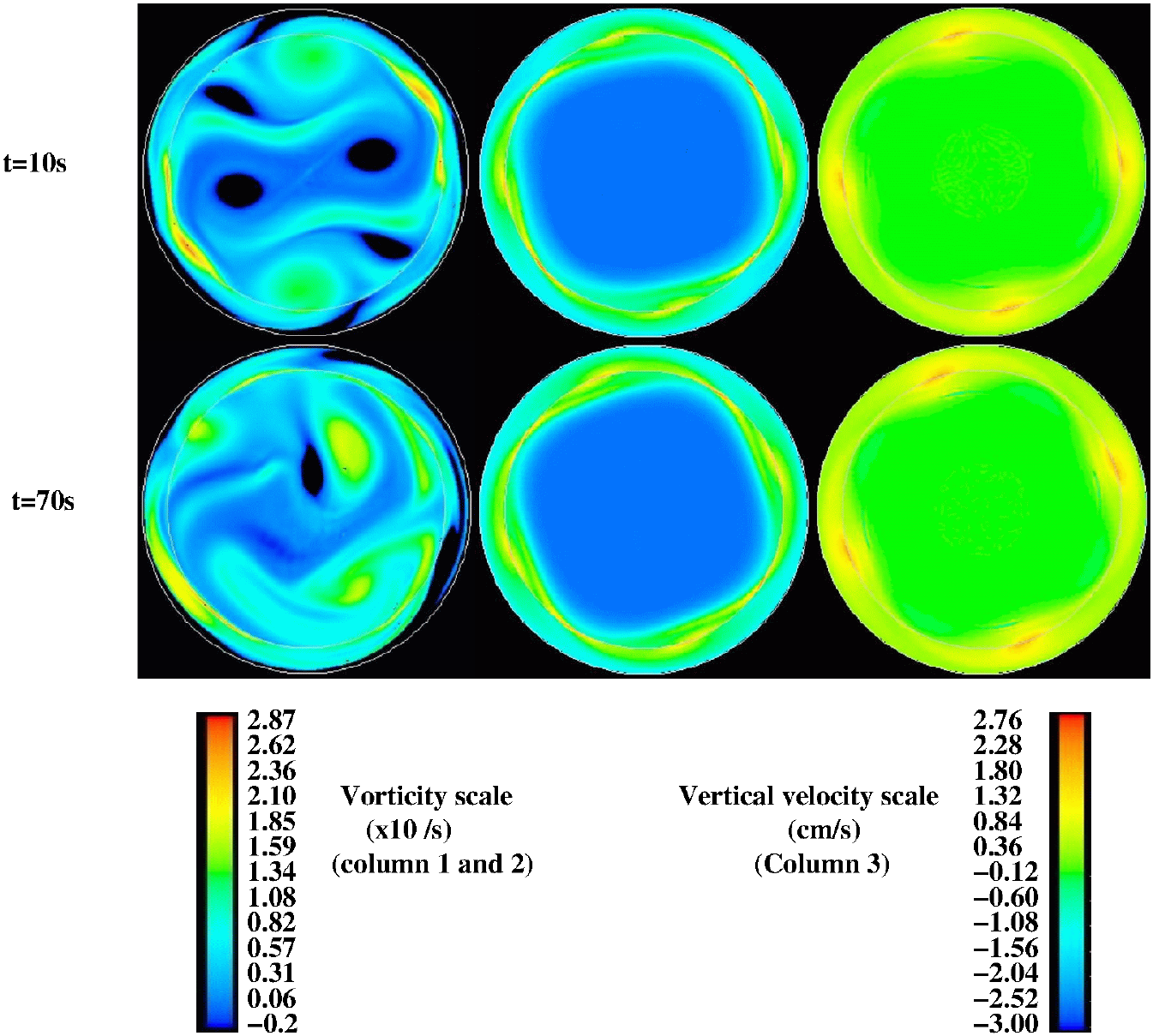}
\centering
\caption{Evolution of the flow with time from numerical simulations
for $I=30$ A, and $B=0.17$ T. Left column:
 vorticity fields obtained using SM82, central column: vorticity
fields obtained using PSM2000. Right Column: vertical velocity fields at
the edge of the Hartmann layer, computed from the horizontal velocity
field given by the simulation of (\ref{PSM2000}).}
\label{films30a_sm_psm}
\end{figure}

\begin{figure}
\centering
\includegraphics[scale=1]{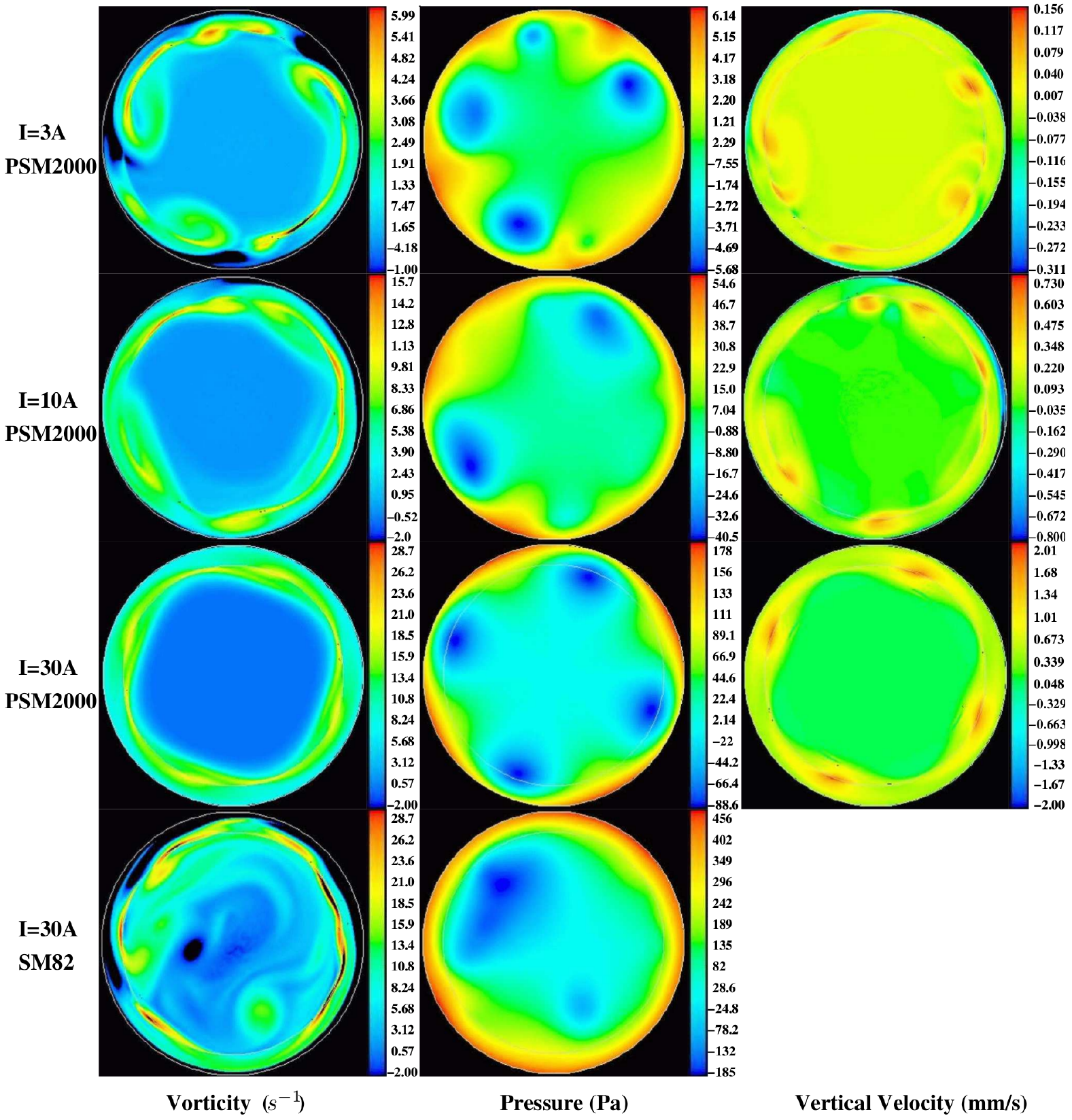}
\caption{Quasi-steady
states of the flow for $B=0.17$ T for different values of
the injected current, obtained from numerical simulations. Left column : vorticity field, central column :
pressure field. Right column : vertical velocity field at the edge of the
Hartmann layer. The dark areas surrounded by blue colour in the vorticity field 
represent negative vorticity (off colorscale). Separation of the boundary layer located at the side 
wall surrounding the flow clearly appears for $I=3$ A (PSM2000) and $I=30$  A (SM82).}
\label{quasi stat}
\end{figure}

The electric current is injected at $t=0$ and remains constant during the
whole simulation. After a few seconds, the azimuthal
velocity of the external annulus reaches the critical value that destabilises
the circular free shear layer located at $r_{e}=0.093m$. This 
Kelvin-Helmholtz instability then produces small cyclonic
vortices, merging into bigger ones (see figure \ref{films30a_sm_psm}).\\
\begin{figure}
\centering
\includegraphics[width=\textwidth]{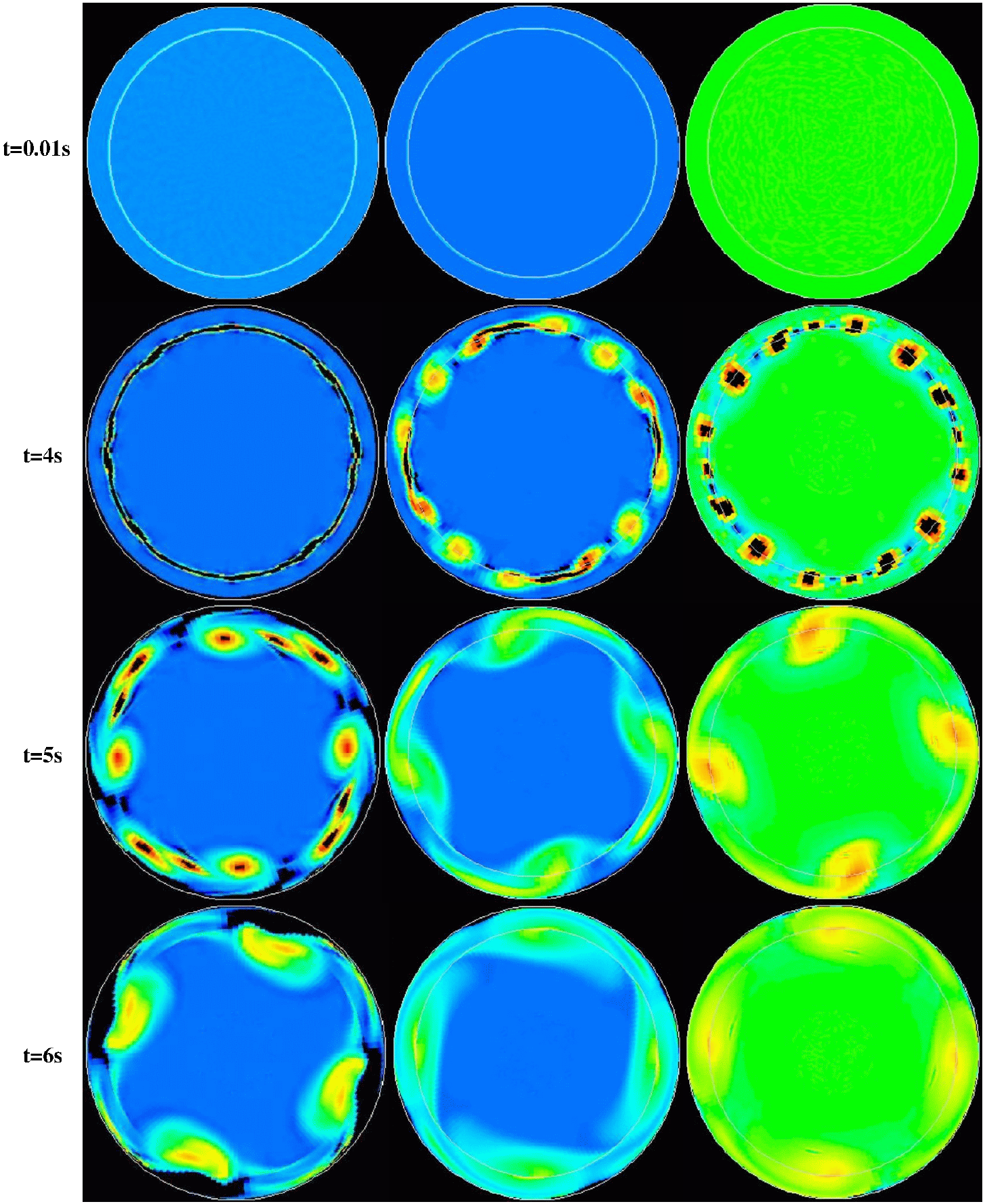}
\caption{See caption on figure  \ref{films30a_sm_psm}.}
\end{figure}
For low injected currents (a few Amp\`{e}res), SM82 and PSM2000 predict flows
 which are very close to each other, 
 but for higher values of $I$, the rotation becomes faster and
Ekman pumping becomes important. As a first
effect, the vorticity structures elongate in the direction of the mean flow
(see figure \ref{quasi stat}$, I=30$ A) in the simulations of 
the PSM2000 model. Notice that this effect
does not affect the pressure field directly. When the phenomenon is strong
enough, vortices cannot move within the rotating reference frame anymore, so
that the final state predicted by the PSM2000 model is made of a few
azimuthally elongated vortices nearly in solid body rotation. For the same current,
the SM82 model predicts a higher rotation speed and a much more chaotic flow,
involving circular vortices of different sizes merging into one another.
According to the results obtained using SM82, boundary layer separations 
also appear for $I=30$ A, at the 
side wall, which lead to the injection of big anticyclonic vortices in the flow 
(These vortices appear as black patches in the pictures of the left column in figures 
\ref{films30a_sm_psm} and \ref{quasi stat}). The
lifetime of such vortices is of the order of magnitude of the inertial time. 
This first view
indicates clearly that the smoothing property of the PSM2000 model shown in
section 2.3.3 can drastically stabilise the flow, to the point 
of literally suppressing turbulence. We shall now
examine the results more quantitatively.
\subsection{Mean velocity profiles}
\subsubsection{Core flow}
\begin{figure}
\centering
\includegraphics[scale=.6]{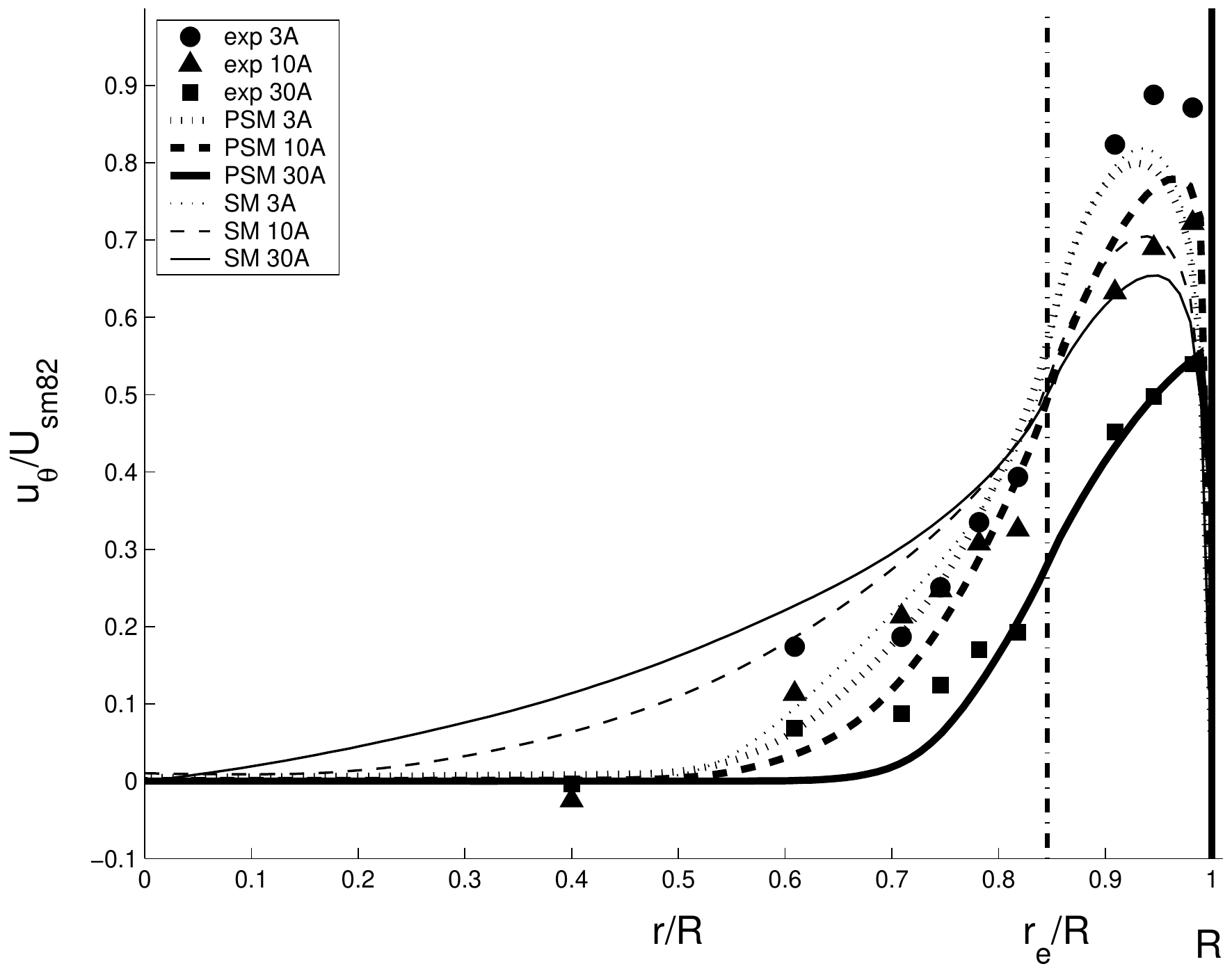}
\caption{Azimuthal velocity profiles at 
quasi steady state for $B=0.17$ T (averaged  on time at quasi-steady state), for several values of the 
injected current.}
\label{profilsV}
\end{figure}

Figure \ref{profilsV} shows the radial profiles of the RMS of the
azimuthal velocity obtained by numerical simulations based on the SM82
and PSM2000 models and by the experiment of \cite{albouss99} respectively. SM82 overestimates 
 the velocity as soon as $I$ reaches
approximately $20$ A, whereas PSM2000 remains in fairly good agreement with
experimental results. The latter however slightly underestimates the
velocity in the inner annulus, near the injection electrodes at $r_{e}$. As
a consequence, the inner half of the free shear layer is a bit thinner than in
the experiments. A more crucial difference is that the SM82 model predicts
a wall side layer of thickness $aHa^{-1/2}$ (which corresponds to the
linear parallel layer theory. See for instance \cite{moreau90}), and which
therefore does not depend on $I$, whereas the radial outward angular momentum
transport associated to secondary flows, squeezes the wall side layer
dramatically in the results obtained using the PSM2000 model (see section \ref{sec:para_lay}).
\subsubsection{Squeezed wall side layers}
\label{sec:shrunk_layers}
\begin{figure}
\centering
\includegraphics[scale=0.7]{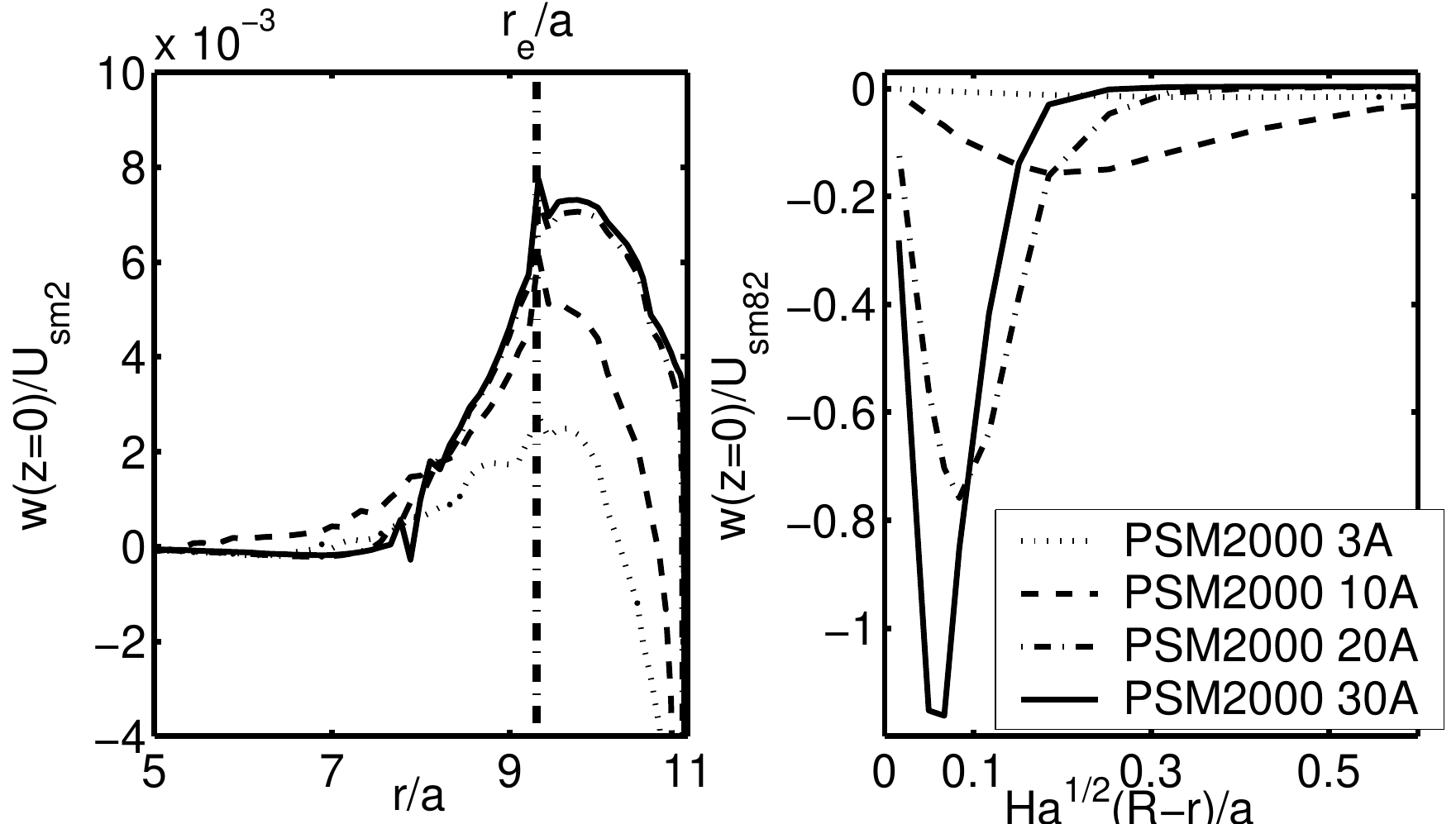}
\caption{Time average at quasi-steady state of the radial profiles of vertical velocity at the edge of the Hartmann
layer, on a whole radius (left: radial position is normalised by $a$) and within
a modified wall-side layer (right: velocities are plotted versus the distance to 
the side wall normalised by the parallel layer thickness in linear regime).}
\label{profilsu3}
\end{figure}

\begin{figure}
\centering
\includegraphics[scale=0.6]{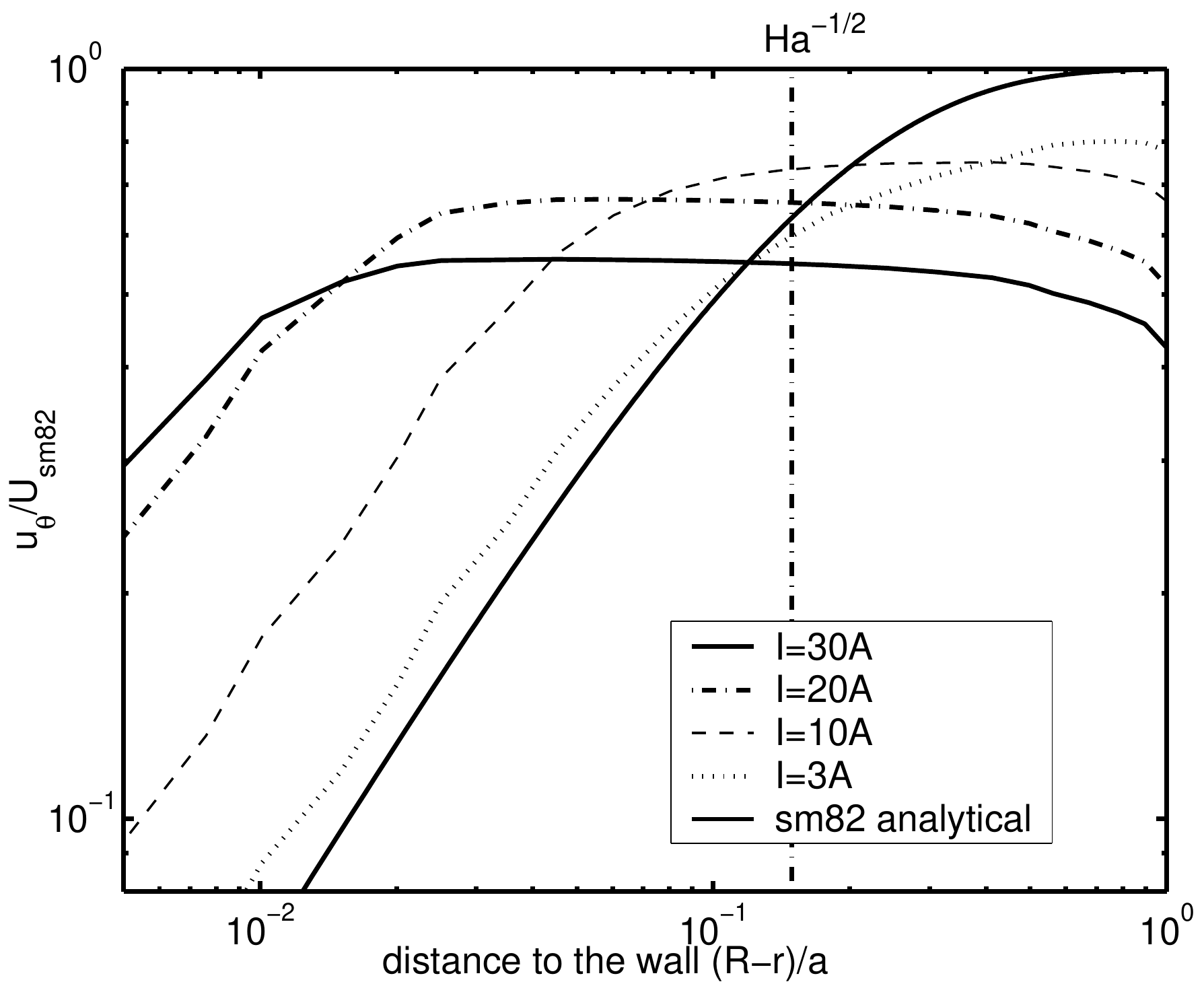}
\caption{Radial profiles of azimuthal velocity within the wall side layer
(averaged in time at quasi-steady state). Velocities are normalised by £
$U_{sm82}$. This graph clearly shows that the higher the forcing, the thinner the boundary layer and the more the maximum velocity is reduced, compared to
flows where recirculations are neglected (which would have a maximum velocity 
closer to $U_{sm82}$). The smooth solid line represents the exponential profile obtained 
analytically from the SM82 model under the assumption of axisymmetry (typical thickness $Ha^{-1/2}$). Remark: The actual numerical computations always rely on
at least 10 points to resolve the wall side boundary layer. The apparent lack 
of precision on the graph is a simple post-processing issue.}
\label{profilsc}
\end{figure}

The vertical velocity at the interface between the Hartmann layer and the core 
(for a mathematically rigorous definition of this interface, see \cite{psm02})
is computed from the solution of the numerical simulation, using the expression
for $w(z=0)$ provided by the PSM2000 model (see \cite{psm00}):
 $w(z=0)=-(5/6) (a^3/\nu) (1/Ha^3)\mathbf{\nabla }_{\bot }\mathbf{.}\left[ \left( \mathbf{\bar u}_{\bot }\mathbf{%
.\nabla }_{\bot }\right)  \mathbf{ \bar u}_{\bot }\right]$.
Figure \ref{profilsu3} shows that a
strong Ekman pumping occurs in the rotating annulus ($r_{e}<r<R$). The small
oscillation appearing in the profile at $r\simeq r_{e}$ indicates that each
big vortex conveyed by the mean flow is subject to a small Ekman pumping
which is added to the global recirculation. This induces an additional radial 
flow. As
the vertical velocity is oriented toward the core in the whole flow, mass
conservation is satisfied thanks to a strong vertical jet occurring at
the wall side layer, the latter being indeed the only area in the flow where $
w(z=0)\leq 0$.\\
Also, the boundary layer at $r=R$ is squeezed by recirculations as 
shown in figure \ref{profilsc}. The mechanisms which explains it is the same as for
isolated vortices described in section \ref{sec:decay}.
%
%

As a  consequence, the flow injected in the core outside of the
wall side layer loops back to the Hartmann layers on a reduced horizontal
area. This makes the already high vertical velocity maximum (oriented toward
the Hartmann layers) in the wall side layer even higher, as shown on 
figure \ref{profilsu3}.

Figure \ref{profilsc} shows the dramatic thinning of the side boundary layer.
Under the assumption of axisymmetry, the PSM2000 model predicts a thickness
of $(36/7)(N/Ha)Rn^{3/2}$ (here $n=2$) for the  parallel layer at the side
wall (see \cite{psm00}, section 4),
which is far thinner than the $aHa^{-1/2}$ thickness of linear parallel
layers. This result however doesn't apply directly here, as it ignores the 
extra recirculations induced by local vortices mentioned in this section.
It is however noteworthy that the modified layer keeps an exponential shape, 
as assumed by \cite{psm00} in order to derive the layer thickness in the 
axisymmetric case.
We shall now see that the phenomenon of wall side layer thinning is much 
more significant in MATUR than in the case of Sommeria's vortices 
(see section \ref{sec:sommeria})
as it reaches a point where it significantly alters the global dissipation.
\subsection{Effect of the secondary flows on global quantities}
\subsubsection{Quasi-steady state}
The direct consequence of the wall side layer being squeezed is that 
velocity gradients are strongly increased near the wall and so is 
the local shear stress.
A good global description of this
effect is provided by the balance of the total angular momentum 
(denoted $L(t)$, and $L_\infty$ at quasi-equilibrium). Hence, we shall 
now investigate how both transients and asymptotic values are affected by 
local and global recirculations.\\

\begin{figure}
\centering
\includegraphics[scale=0.6]{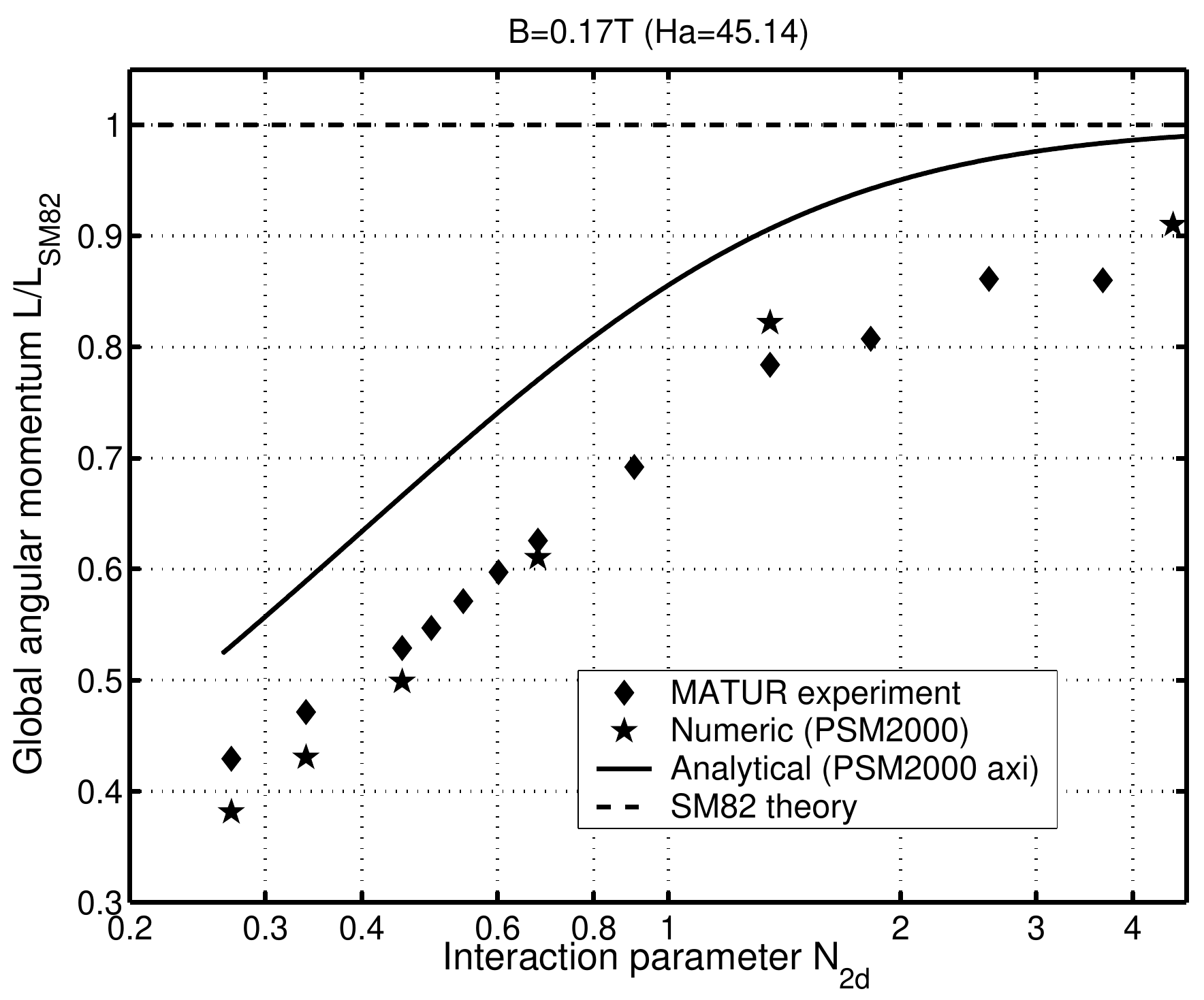}
\caption{Global angular momentum for $B=0.17$ T, at quasi-equilibrium  as a 
function of the interaction parameter. For low values of $N_{2d}$, the global
 angular momentum is reduced 
 to about a third of its value without non-linear effects (this corresponds to 
 a drop in the maximal azimuthal velocity to about half the SM82 value). 
 Note that all values 
 are normalised by the theoretical value at equilibrium derived from the 
 SM82 model $L_{sm82}$.}
\label{lb017}
\end{figure}

\begin{figure}
\centering
\includegraphics[scale=0.4,angle=-90]{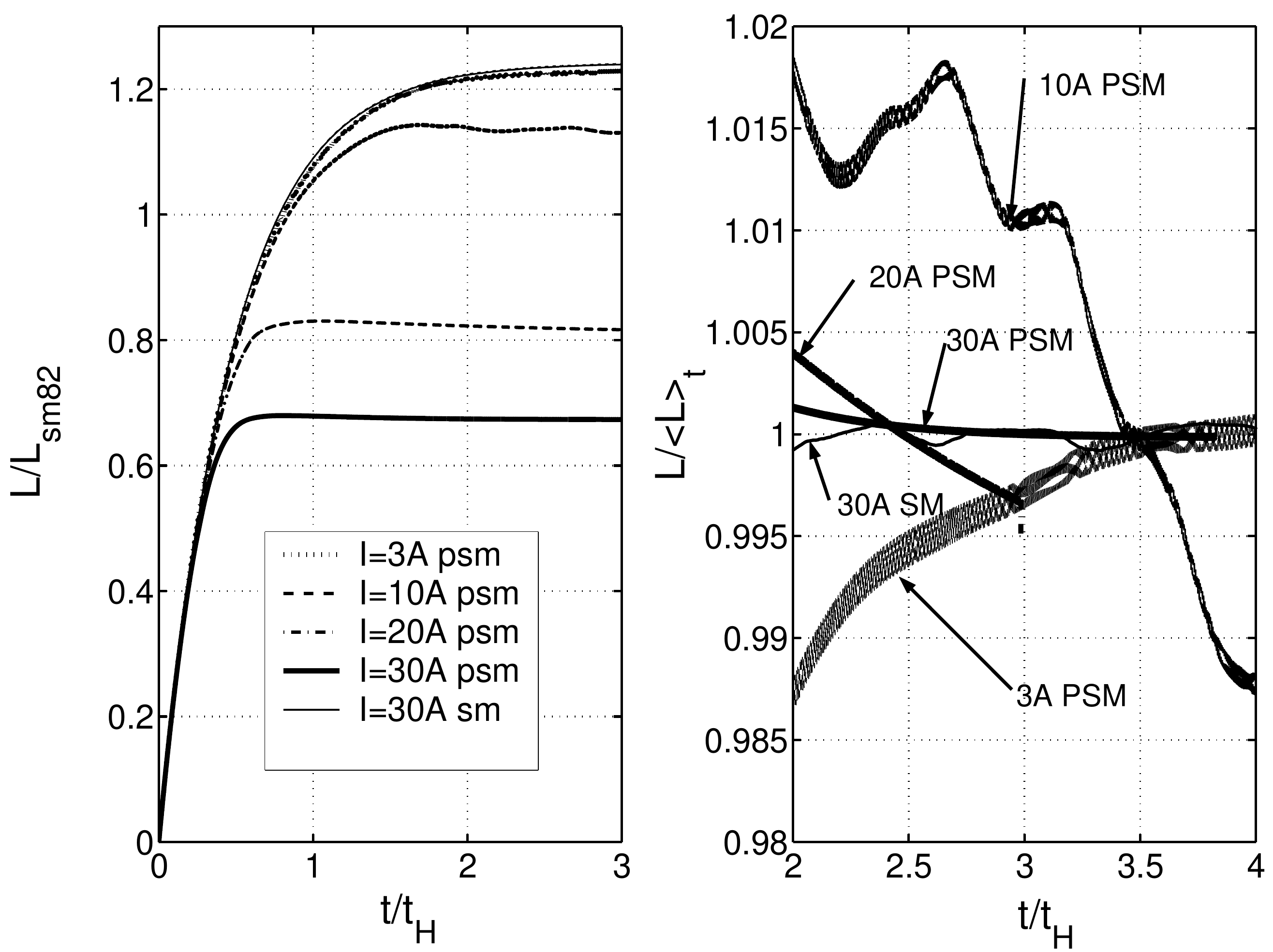}
\caption{Evolution of the global angular momentum with time.
On the left figure, curves $I=3$ A (PSM2000) and $I=30$ A (SM) nearly merge as 
recirculations are very weak and don't affect the flow very much for $I=3$ A.
On the right figure, quantities are normalised by their value at 
quasi-equilibrium, averaged over a period of $t_H$. The thickness of the lines
give an idea of the numerical precision of the result (about $0.1\%$)}
\label{int_l}
\end{figure}

Figure \ref{lb017} shows a comparison between the global angular momentum at 
quasi-equilibrium
$L_\infty$ measured in the experiment, analytical results (derived from SM82 and PSM2000
 with the assumption of axisymmetry by \cite{psm00}), and our numerical simulations. 
The  theoretical value of PSM2000
is about $15 \%$ away from the
experimental results whereas the full simulation of (\ref{PSM2000})
gives a far more accurate result. The
main difference between the two models is the axisymmetry assumption
: in the full simulation, the recirculation associated to cyclonic vortices
causes dissipation in the wall side layer so that the flow is slightly more
damped than in the axisymmetric case.
Another important effect of recirculations is the
"stabilisation" of the flow. Indeed, figure \ref{int_l} 
shows that
the amplitude of the oscillations of the global angular momentum  at 
equilibrium is strongly
reduced, compared to SM82 results, which corresponds to the observation that
the flow is less chaotic when significant recirculations occur. 
Global enstrophy (\textit{resp.} energy) oscillates by around $5 \%$ 
(\textit{resp.} $10\%$) at $3$ A with PSM2000 and $30$ A 
with SM82. 
This oscillation falls below $0.1 \%$ (\textit{resp.} $1\%$) for $I=30$ A with PSM2000. 

This is a
consequence of the local damping of disturbances pointed out in section
\ref{sec:vortex}, which does not appear in SM82 simulations.
%
%

\subsubsection{Transient time}
\label{sec:trans}
\begin{figure}
\centering
\includegraphics[scale=0.6]{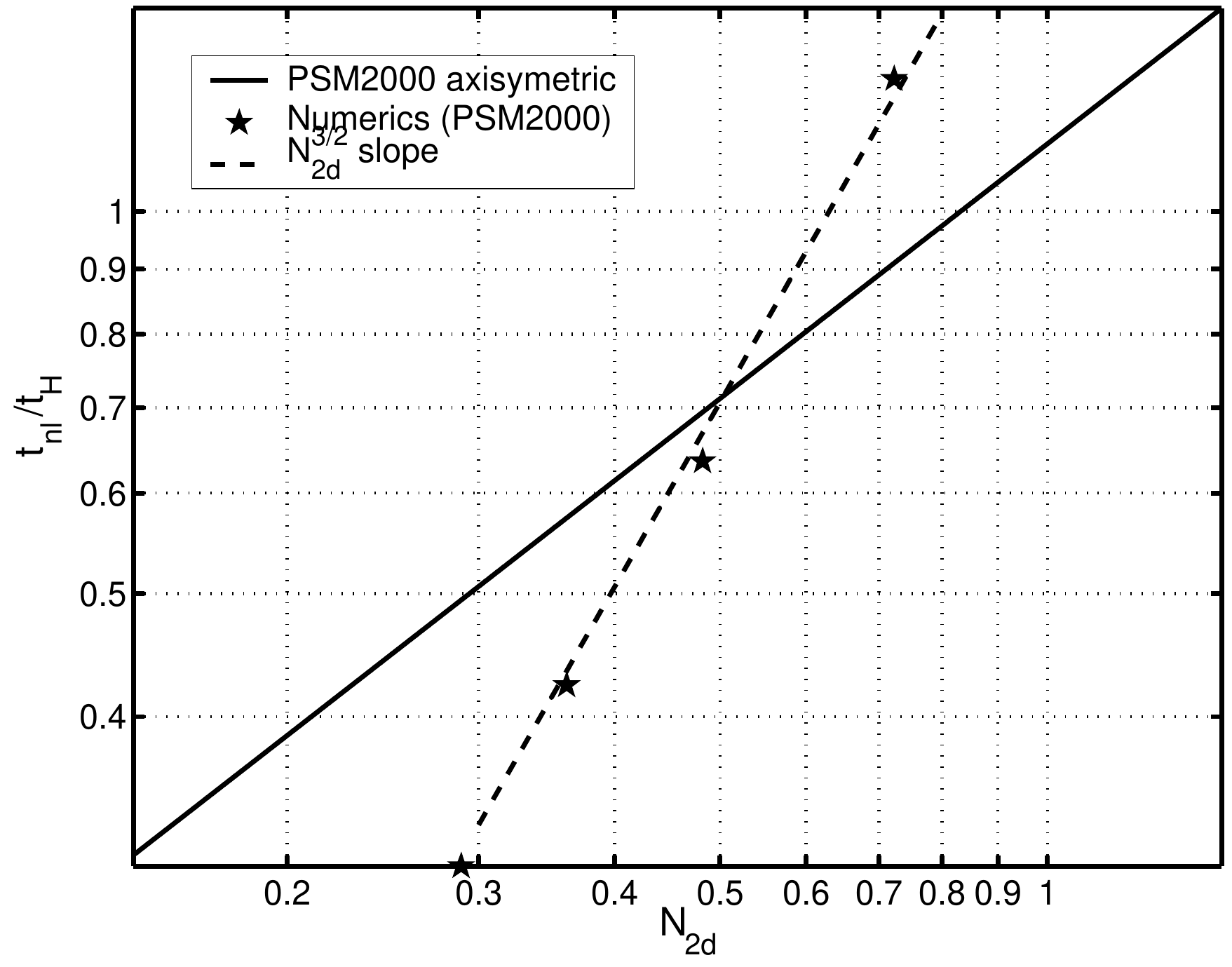}
\caption{Transient time obtained numerically after switching on the 
forcing on a fluid 
at rest (stars) and non-linear time $t_{nl}$ derived from PSM2000 under 
the assumption of axisymmetry, versus total interaction parameter.}
\label{transient}
\end{figure}

 The SM82 model, predicts that the system should reach the quasi-steady state
in a time of the order of $t_{H}$. When important 
Ekman pumping occurs, the wall side layers can become thin enough
to significantly increase the global dissipation. This results in 
shortened response time of the flow. This 
tendency is illustrated in Figure \ref{transient} which shows 
that the typical response time of the flow near quasi-equilibrium varies
 approximately as $N^{3/2}$ (in practise, this time is obtained by measuring the slope
 of the $L(t)-L(t\rightarrow \infty)$ curve near equilibrium in a log-log 
diagram). Using the axisymmetric assumption, the evolution equation for the 
angular momentum  
derived from the PSM2000 model can be linearised around the quasi-steady 
state: this provides a response time varying as $N^{2/3}Ha^{1/3}$. The reason for the 
 difference is again that the full numerical simulation accounts for local recirculations, added to the recirculation due to global rotation by each vortex. As discussed in 
section \ref{sec:shrunk_layers}, these additional recirculations make the wall 
 side layer even thinner and increase the wall shear stress compared to the axisymmetric 
 case.
\subsection{Stability of the free shear layer}
\label{sec:para_lay}
The axisymmetric free shear layer located at $r=r_{e}$ is subject to a 
Kelvin-Helmholtz instability, which leads to the growth of cyclonic vortices 
along
the layer. In order to get a
rough estimate of the stability condition, the radial profile of azimuthal
velocity is assumed linear and the layer is assumed to be of thickness $%
\delta =aHa^{-1/2}.$ Without viscous or electromagnetic effects, the layer is unconditionally
unstable. The most unstable radial wavenumber $k$ and the related growth
rate $\sigma _{c}$ are given by (see \cite{chandra61}): 
\begin{subequations}
\begin{gather}
k\delta =0.8  \label{k unstable} \\
\sigma _{c}=0.4\dfrac{U}{\delta }  \label{sig unstable}
\end{gather}
\end{subequations}
In the MHD problem, the magnetic field tends to 
stabilise the flow because of the Hartmann friction. Indeed, for small
enough velocities (experimentally, theses velocities correspond to values of 
$I$ below $0.2$ A), the laminar parallel layer can be stable if $t_{H}^{-1}$
is bigger than the frictionless growth rate : 
\begin{equation}
0.4\frac{U}{a}Ha^{-\frac{1}{2}}<\frac{1}{t_H}
\end{equation}%
or equivalently, using the Reynolds number $Re=Ua/\nu$: 
\begin{equation}
\frac{Re}{Ha^{\frac{1}{2}}}<2.5
\label{stab. marginale}
\end{equation}
In other words, the piecewise linear profile chosen for the parallel layer becomes 
linearly unstable when the Reynolds number built on its thickness exceeds the 
threshold of $2.5$.
At $B=0.17$ T , the typical size of the vortices appearing at the onset of
instability is given by (\ref{k unstable}) and corresponds to $2.6$ mm.
\cite{lieutaud01} have performed an energetic stability study of the 
2d problem, using a more realistic piecewise exponential profile. They find a
stability threshold (below which any arbitrary perturbation is damped) 
$Re/\sqrt{Ha}=9$ and a most unstable wavelength of $2$ mm. The fact
 that even under slightly different assumptions, both linear and energetic 
 stability threshold remain of comparable orders of magnitude suggests that 
 the free shear layer is indeed destabilised by infinitesimal perturbations 
 of typical wavelength close to the boundary layer thickness.\\
 
 Finally, it is worth mentioning the effect of curvature: \cite{liou94} has 
 shown that for stably curved layers (\textit{i.e.} high speed stream on the 
 outside of the curvature) the centrifugal force tends to slightly reduce the growth rate of 
 the Kelvin-Helmholtz unstable modes, which might increase the instability 
 threshold, 
 without affecting the basic mechanism.\\

In both numerical simulations and experiment, The destabilised state is
itself unstable and the vortices merge until a small
number of big structures is reached. The choice of either SM82 or PSM2000
does not affect significantly the instability found by numerical simulations.
 Actually, PSM2000 leads to an
earlier destabilisation ($t=5$ s at $30$ A versus $t=4$ s for the SM82 model),
but this is due to the fact that non-linear effects tend to reduce the
characteristic response time of the flow (see section \ref{sec:trans}) so
that the unstable regime is reached quicker with PSM2000.
In the numerical simulations, the laminar free shear layer is only radially
discretised with two or three points as explained earlier, so that the 
numerical profile is rather close
to the piecewise linear profile studied in this section.
\subsection{2d fluctuations}
\begin{figure}
\centering
\includegraphics[scale=0.5]{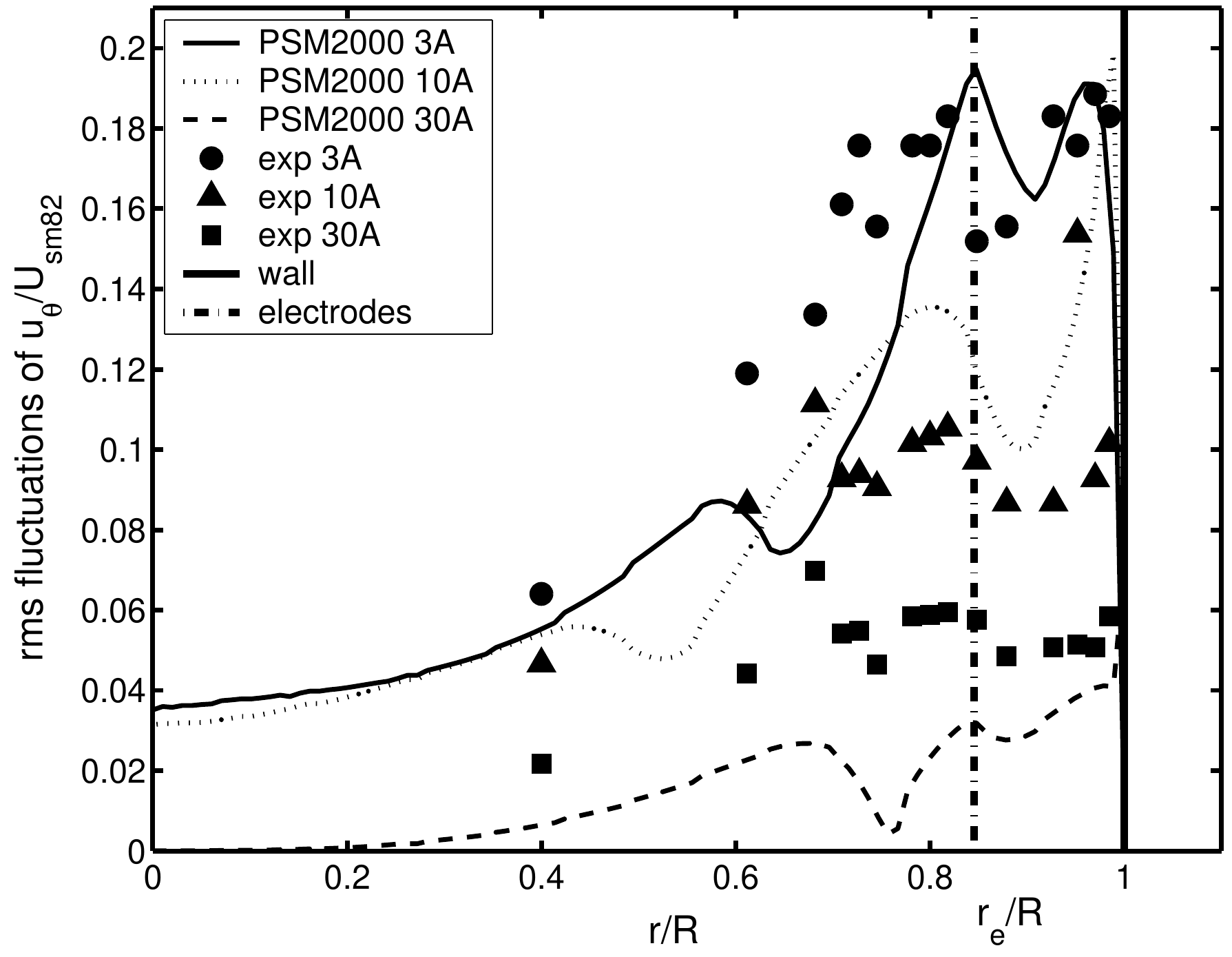}
\includegraphics[scale=0.5]{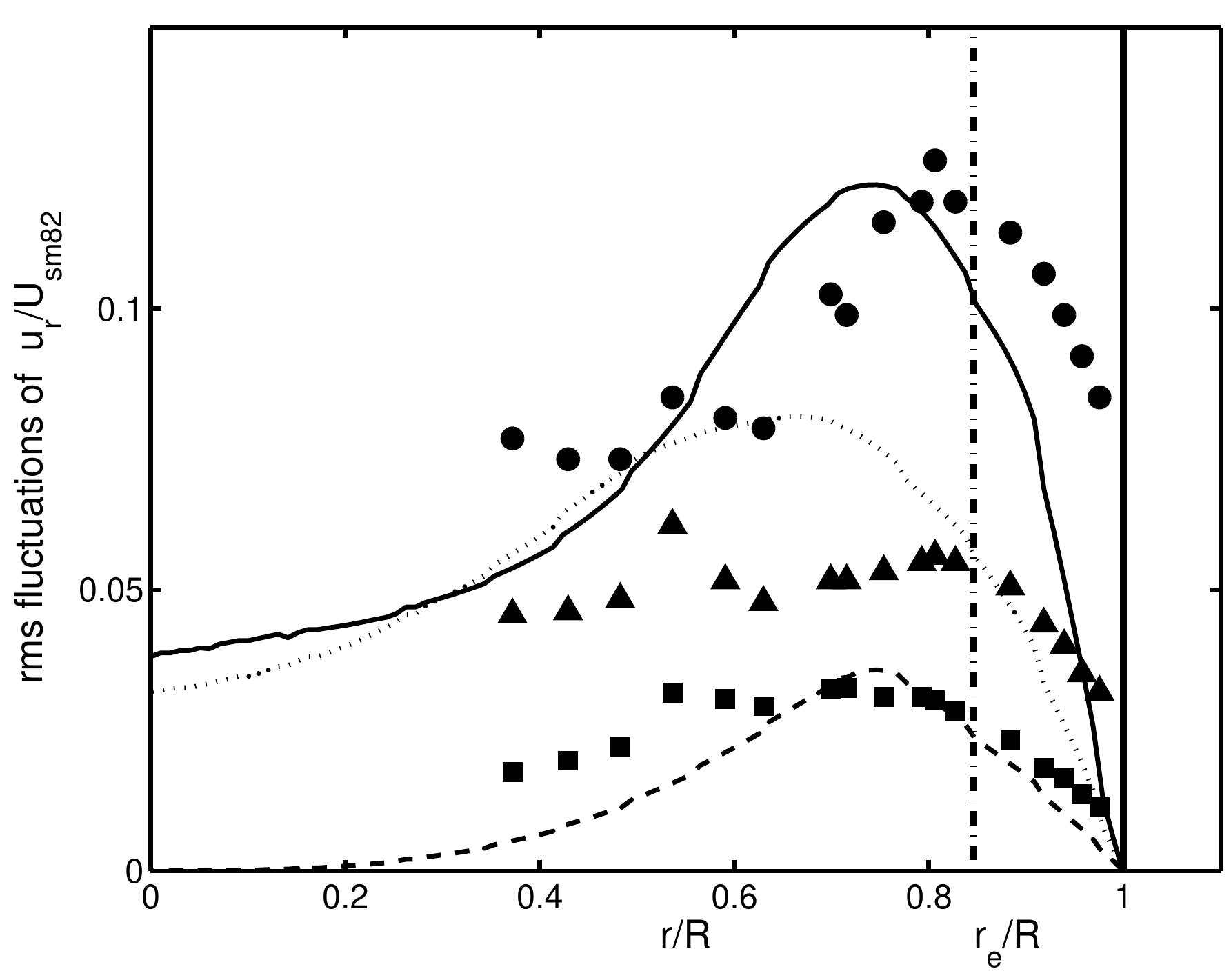}                                          
\caption{Radial profiles of Root Mean Square azimuthal (top) and radial (bottom)
velocity fluctuations (averaged in time at quasi-steady state), for several values of the 
injected current.}
\label{rms}
\end{figure}

Radial profiles of RMS azimuthal velocity fluctuations are in
good agreement with experimental measurements (see figure 
 \ref{rms}): both exhibit two extrema at $r=0.07$ m and $r=r_{e}$.
The area $r=0.07$ m corresponds to the location inside the electrodes ring where
 the average velocity is very low but perturbed by the edge of passing vortices, which explains the important fluctuations of velocity.
It also clearly appears that the relative intensity of the velocity fluctuations decreases with 
decreasing $N$, \text{i.e.} when secondary flows become stronger. 
This phenomenon is more than 
likely related to the smoothing effect theoretically predicted in section \ref{sec:vortex}
 and visible in figures \ref{films30a_sm_psm} and \ref{quasi stat}. 
It should also be noticed that the velocity fluctuations can be of the 
order of $1$ cm/s or less. At such low velocities, the experimental results are not as precise as 
for higher velocities such as those in figure \ref{profilsc}. The agreement 
between theory and experiment should therefore be considered to be as good as one can expect.\\
\begin{figure}
\centering
\includegraphics[scale=0.6]{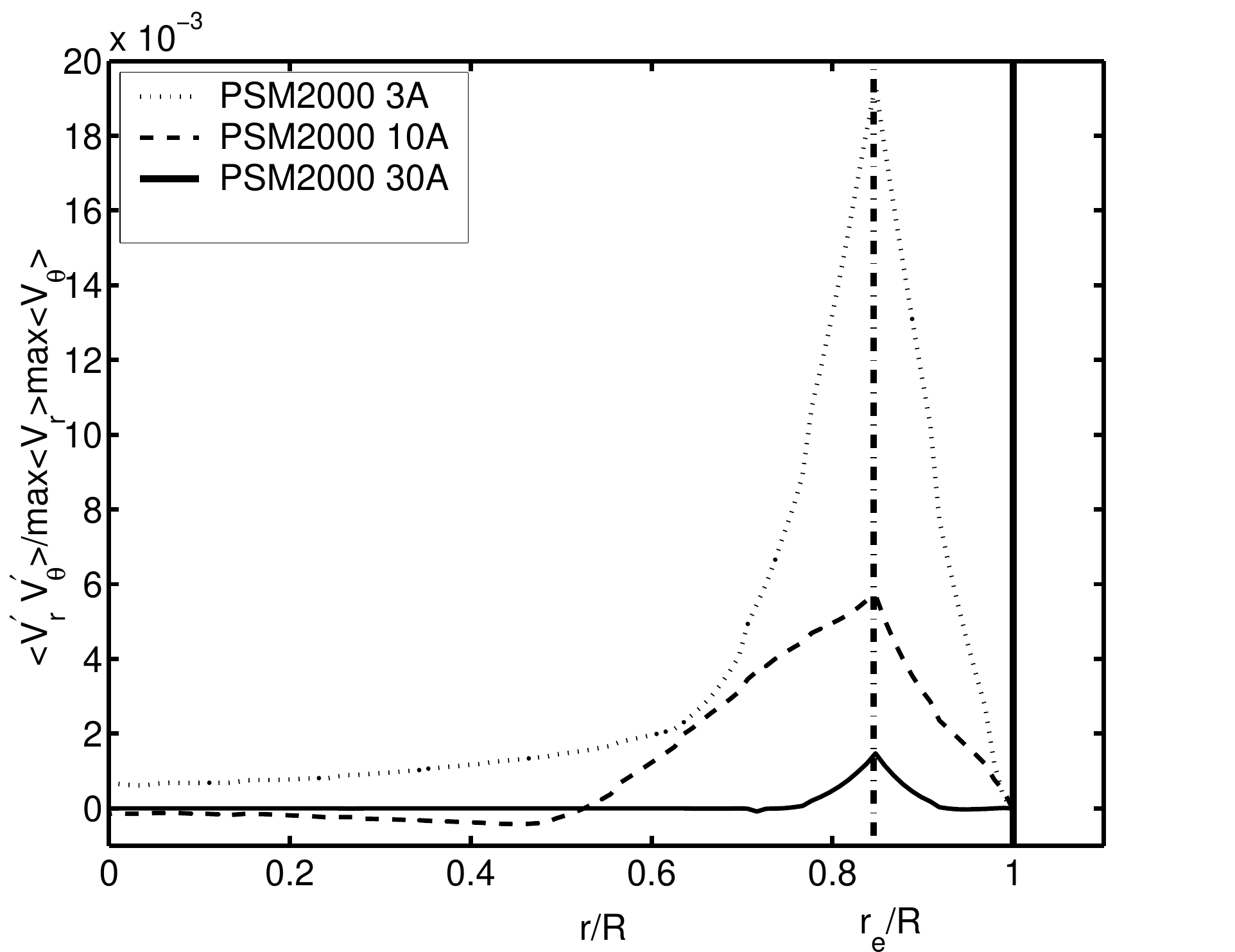}
\caption{Radial profiles of Root Mean Square value of the correlations of  
velocity fluctuations (averaged on time at quasi-steady state), 
for several values of the injected current. Values are normalised by the maximum
values reached in the time-averaged radial velocity 
profiles. These curves therefore give a measure of the turbulent intensity.}
\label{fig:turbu_int}
\end{figure}

The damping of turbulent fluctuations by local Ekman recirculations is 
more visible when looking at the turbulent intensities plotted on 
figure \ref{fig:turbu_int}. It shows the radial profile of 
$<v_r^\prime v_\theta^\prime>$,  the values 
of which are also strongly reduced by Ekman recirculations. But what further 
appears, is  
that for higher forcing, all non-zero values of $<v_r^\prime v_\theta^\prime>$ 
are are confined radially around  the 
injection electrode. For $I=20$ A and $I=30$ A, the intensity of the correlation
decreases almost linearly with the distance to the electrode. In other words,
apart from the fluctuations due to passage of big vortices in almost solid rotation, 
there is hardly any turbulent fluctuations left. Moreover, the typical width 
of the vortices (indicated by the width of the peak in the values of 
$<v_r^\prime v_\theta^\prime>$ ) strongly decreases with increasing forcing.\\

Four distinct  mechanisms dissipate energy in the flow: turbulent dissipation, 
friction in the Hartmann layers, friction in the side wall and local dissipation
by secondary flows. The typical ratio between turbulent dissipation and 
Hartmann damping is around $10^{-3}$, which confirms that turbulent dissipation 
is very small, as expected in 2d turbulence. The dissipation in the side layer 
is drastically increased by the radial transport of angular momentum due to 
Ekman pumping. One can get an idea about the importance of this dissipation 
by comparing the analytic values obtained for the angular momentum at quasi-
equilibrium using SM82 (which ignores the recirculations) and PSM2000 (see
figure \ref{lb017}). For 
$I=30$ A, dissipation in the side layer is of the order of the Hartmann
dissipation. This analytical value is obtained under the assumption of 
axisymmetry and therefore ignores the local dissipation due to local
recirculations. The fact that it doesn't depart significantly from the 
experiment suggests that this local dissipation is rather weak.
\subsection{Higher fields and turbulent Hartmann layer}
For higher magnetic fields ($B=0.5$ T) and strong forcing ($I=30$ A), the ratio
$Ha/N$ becomes large ($282$). This ratio also represents the Reynolds number
scaled on the thickness of the Hartmann layer and it is well known that the
Hartmann layer becomes turbulent when it reaches such values ($250$ 
according to the experimental study of \cite{hua74}, $380$ according to the 
experiments of \cite{moresco04} and 390 according to the numerical work of
\cite{krasnov04}, see also the theoretical work by \cite{albouss00}). 
\cite{krasnov04} also found that even when the Hartmann layer
becomes turbulent, the core flow can still remain 2d. It will 
indeed be the case if the turnover time associated with 3d velocity 
fluctuations,  
remains smaller than the typical 
bidimensionalisation time. \cite{sm82} have shown that if $k$ is the 
non-dimensional wavenumber (normalised by $1/a$) associated with one particular 
structure, this structure is two-dimensional if $k<<N^{-1/3}$, which can be 
satisfied over the whole spectrum of $k$ even for values of $Ha/N$ above the 
Hartmann layer stability threshold.\\

\begin{figure}
\centering
\includegraphics[scale=0.5]{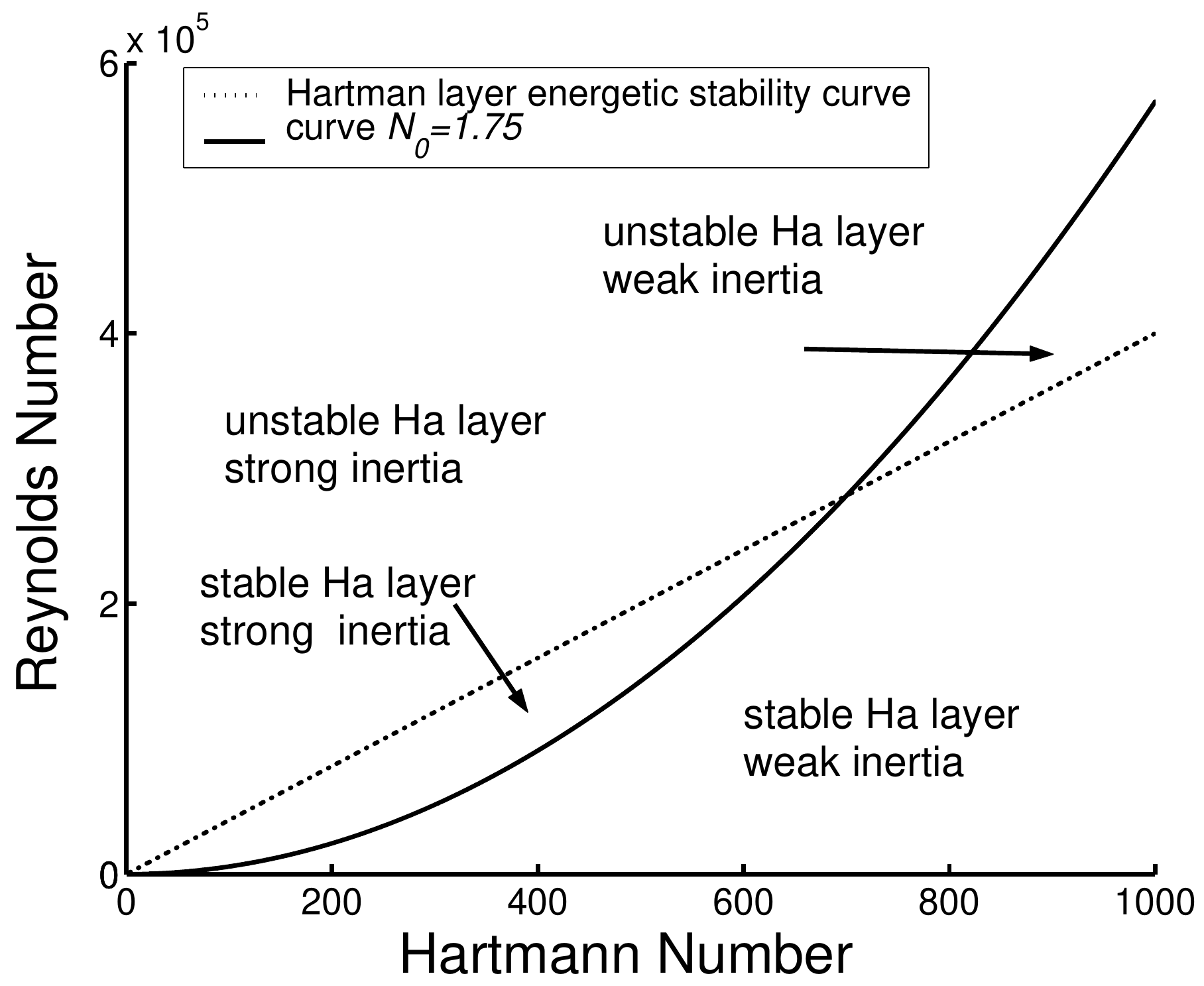}
\caption{Hartmann layer stability and inertial effects diagram. The stability 
threshold is arbitrarily fixed at $Re/Ha=400$.
The curve of constant amplitude inertial effects is plotted for $N=1.75$.}
\label{hadiag}
\end{figure}

For such high values of $Ha/N$, the global angular momentum computed 
from the
PSM2000 model exhibits a strong discrepancy with experimental results. 
The reason is that the Hartmann layer becomes
turbulent. Indeed, the magnitude of non-linear effects due to Ekman
recirculation is monitored by the interaction parameter $N\sim B^2/U$, 
which means that if the magnetic field is increased, the velocity has
to increase as $B^{2}$ to observe non-linear effects of the same
magnitude. The Hartmann layer
becomes turbulent when the Reynolds number at the scale of the layer 
${Re}/Ha\sim U/B \sim B/N$ exceeds a few hundred. 
For a fixed value
of $N$ (\textit{i.e}. given relative recirculation magnitude), this
threshold is then lower for lower fields. In other words, for sufficiently high
magnetics fields, the Hartmann layer is already turbulent when values of $U$
are reached, which are high enough to induce a significant Ekman pumping. 
Both SM82
and PSM2000 models rely on the assumption that the Hartmann layer is laminar and therefore 
cannot represent the flow above the Hartmann layer stability threshold
(see figure \ref{hadiag}).
%
%
\section{Conclusion}
The comparison between the predictions derived from the PSM2000 model and
 the experimental 
results of \cite{albouss99} shows that the model achieves a good accuracy
 for all measured quantities, and this in spite of its relative simplicity. 
The effects of both local (at the scale of large eddies) and global (at the scale 
of the whole cell) recirculations are reproduced in a fairly realistic way. 
Moreover, the new model allows us to point out quite simply the 3d details of their 
mechanisms, whilst retaining the simplicity of 2d calculations.
It is worth mentioning here, three of the major properties of this model. 
Firstly, the 
second-order non-linear terms yield a tendency to smooth the velocity gradients, 
which can ultimately erase the chaotic behaviour of the flow and damp 2d 
turbulence. Secondly, they induce 
some additional dissipation within the Parallel boundary layers in which the 
velocity gradients are increased. Finally, it 
appears that the response time of the flow is reduced. 
The latter effect seems to be related to the transport of any quantity by the secondary 
flows. Broadly speaking, the quasi-2d turbulent flow tends to be more homogeneous.\\

We now wish to mention two questions, which remain open. Firstly, the secondary centrifugal flows 
which characterise PSM2000 should certainly affect the transport of any passive scalar quantity. 
This might be investigated by adding an energy equation to (\ref{PSM2000}) and the accuracy of the results 
might be checked by comparison with the temperature measurements of \cite{albouss99}. 
Second, both SM82 and PSM2000 fail to model the turbulence within the Hartmann layer when it 
is present. Its consequence should be to increase the layer's thickness and the wall friction. 
A new MHD 2d model could be derived from the model by \cite{albouss00}  
for the turbulent Hartmann layer.\\

Finally, we insist that both examples of the SM82 and PSM2000 models do not only offer a method, but 
also prove
that this method is flexible enough to  make the modelling of complex 
3d flows possible, as long as there is a local model for the phenomenon involved(here we combine MHD and rotation effect).

When applicable, this 
appears to the authors as a good alternative to fully-3d simulations which require enormous
computational resources. This is all the more important as 3d CFD is sometimes 
only possible at the 
expense of rather unphysical approximations or numerical adjustments. Unlike these, 
PSM2000-like models are rigorously derived from the equations thanks to well controlled
 approximations, which ensure the reliability and clearly mark their area of validity.
 Thanks to these features, the refined 2d model has proven accurate enough to 
 point out a property which had not been mentioned before to our knowledge: the
 non-linear smoothing by local recirculations.\\

	This method can also be extended to any kind of quasi-2d flow, such as rotating flows. The analogy between the kind of flow described in 
this paper and some geophysical flows (see \cite{Poth00}) suggests that 
corrections such as those featured in  PSM2000 could turn out to be efficient 
in 
modelling oceans or atmospheres. \cite{dellar03} has indeed recently shown that
PSM2000 exhibits a very similar behaviour to the model developed by 
\cite{benzi90} for 2d turbulence.\\

The authors are particularly grateful to Martin Cowley for his active contribution 
to the presentation of the 2d models, as well as to the discussions around the meaning
of these models.


~                                                                                                                                                                                                                                           
~                                                                                                                                                                                                                                           
~

\end{document}